\documentclass[prd,aps,onecolumn,showpacs, nofootinbib,superscriptaddress,notitlepage]{revtex4-1}
\usepackage{amsmath,graphicx,epsfig,amssymb,amsthm,bm,mathrsfs}
\usepackage{adjustbox}
\usepackage{array}
\usepackage{inputenc}
\usepackage[usenames]{color}
\usepackage{slashed}
\usepackage{multirow}
\usepackage{subfigure}
\usepackage{dsfont}
\usepackage{comment}
\usepackage{steinmetz}
\usepackage{setspace}
\usepackage{simpler-wick}
\usepackage{caption}

\allowdisplaybreaks

\usepackage{floatrow} 
\newfloatcommand{capbtabbox}{table}[][\FBwidth]

\begin{document}
\title{Lightcone and quasi distribution amplitudes for light octet and decuplet baryons}

\author{Chao Han}
\affiliation{INPAC, Key Laboratory for Particle Astrophysics and Cosmology (MOE),  Shanghai Key Laboratory for Particle Physics and Cosmology, School of Physics and Astronomy, Shanghai Jiao Tong University, Shanghai 200240, China}
\author{Wei Wang} \email{Corresponding author: wei.wang@sjtu.edu.cn}
\affiliation{INPAC, Key Laboratory for Particle Astrophysics and Cosmology (MOE),  Shanghai Key Laboratory for Particle Physics and Cosmology, School of Physics and Astronomy, Shanghai Jiao Tong University, Shanghai 200240, China}
\affiliation{Southern Center for Nuclear-Science Theory (SCNT), Institute of Modern Physics, Chinese Academy of Sciences, Huizhou 516000, Guangdong Province, China}
\author{Jun Zeng}\email{Corresponding author: zengj@sjtu.edu.cn }
\affiliation{INPAC, Key Laboratory for Particle Astrophysics and Cosmology (MOE),  Shanghai Key Laboratory for Particle Physics and Cosmology, School of Physics and Astronomy, Shanghai Jiao Tong University, Shanghai 200240, China}
\author{Jia-Lu Zhang}\email{Corresponding author: elpsycongr00@sjtu.edu.cn }
\affiliation{INPAC, Key Laboratory for Particle Astrophysics and Cosmology (MOE),  Shanghai Key Laboratory for Particle Physics and Cosmology, School of Physics and Astronomy, Shanghai Jiao Tong University, Shanghai 200240, China}

\date{\today}

\begin{abstract} 
We present a comprehensive investigation of leading-twist lightcone distribution amplitudes (LCDAs) and quasi distribution amplitudes (quasi-DAs) for light octet and decuplet baryons within  large momentum effective theory. In LaMET, LCDAs can be factorized in terms of a hard kernel and quasi-DAs that are defined as spatial correlators and calculable   on Lattice QCD. 
To renormalize quasi-DAs and eliminate the singular terms $\ln(\mu^2 z_i^2)'s$ in them, which undermine the efficacy in perturbative expansion, we adopt a hybrid renormalization scheme that combines the self-renormalization and ratio scheme. Through self-renormalization, we eliminate UV divergences and linear divergences at large spatial separations in quasi distribution amplitudes without introducing additional nonperturbative effects. By taking a ratio with respect to the zero-momentum matrix element, we effectively remove UV divergences at small spatial separations. 
Under the hybrid renormalization scheme, we calculate the hard kernels up to one-loop accuracy. It turns out that only four different hard kernels are needed for all leading-twist LCDAs of octet and decuplet baryons.  These results are crucial for the lattice-based exploration of the LCDAs of a light baryon from the first principle of QCD.
\end{abstract}

\maketitle

\section{introduction}

Lightcone distribution amplitudes (LCDAs) of mesons and baryons are fundamental non-perturbative quantities  in QCD that describe how the longitudinal momentum of quarks and gluons is distributed within a hadron. LCDAs, together with parton distribution functions (PDFs), provide a comprehensive depiction of hadron structure. They also serve as a non-perturbative input in exclusive processes with large momentum transfers~\cite{Shih:1998pb}. Besides, the LCDAs of a light baryon can find wide applications in various scenarios, ranging from studying weak decays in bottom baryons to extract the CKM matrix element $|V_{ub}|$~\cite{LHCb:2015eia} to probing new physics beyond the standard model through flavor-changing neutral current processes~\cite{LHCb:2015tgy,LHCb:2021byf,LHCb:2023ptw}.  Examples which utilize the baryon LCDAs to predict the decay rates of heavy baryons based on factorization theorems can be found in Refs.~\cite{Chou:2001bn,Huang:2004vf,Wang:2011uv,Han:2022srw,Khodjamirian:2023wol}.

Despite their crucial importance, the development of baryon LCDAs has not progressed as smoothly as desired.  The extraction of LCDAs from experiments is particularly challenging since it needs to relate baryon LCDAs to experimentally measurable quantities, involving intricate assumptions and simplifications. Thus,  in contrast to PDFs, which have been extensively constrained by experimental data from deep inelastic scattering and other processes (for a recent review, see Ref.~\cite{Gao:2017yyd}), LCDAs suffer from fewer direct experimental constraints.  
On theoretical side, calculating LCDAs from first-principle techniques like Lattice QCD meets computational and technical challenges, caused by the Euclidean nature of the method when applied to lightcone quantities. Among the available analyses on the baryon LCDAs from Lattice QCD~\cite{Gockeler:2008xv,QCDSF:2008qtn,Lenz:2009ar,Bali:2015ykx,RQCD:2019hps}, most are limited to extract only a few lowest moments of the octet baryons. From phenomenological viewpoint, unavoidable  uncertainties in  modeling baryon LCDAs  can arise from assumptions made in the frameworks used to describe them.
QCD sum rules~\cite{Chernyak:1984bm,King:1986wi,Chernyak:1987nu,Farrar:1988vz,Stefanis:1992nw,Braun:1999te,Liu:2008yg,Liu:2009uc,Anikin:2013aka}, which is a prevalent approach to study LCDAs, also contains sizable systematic uncertainties,  not to mention the uncontrollable systematic uncertainties in  other phenomenological models, as discussed in~\cite{Dziembowski:1987zq,Bolz:1996sw,Petrov:2002jr,Pasquini:2009ki,Maji:2016yqo,Kim:2021zbz}.

In a series of works~\cite{Deng:2023csv,Han:2023xbl,Han:2023hgy}, a hadron-to-vacuum spatial correlator is employed to extract the baryon LCDAs, providing a promising step forward in addressing the theoretical challenges. 
This progress is achieved within the framework of large-momentum effective theory (LaMET)~\cite{Ji:2013dva,Ji:2014gla} (see Refs.~\cite{Cichy:2018mum,Zhao:2018fyu,Ji:2020ect} for a review), which establishes a relation between LCDAs and the Fourier transformation of the spatial correlator whose hadron state has a large but finite momentum, referred to as quasi-DAs.
It has been applied to the study of  the LCDAs~\cite{Zhang:2017bzy,Bali:2018spj,Xu:2018mpf,Zhang:2017zfe,Liu:2018tox,Wang:2019msf,Zhang:2020gaj,Hua:2020gnw,LatticeParton:2022zqc,Hu:2023bba}, albeit only focusing on mesons.
When the hadron momentum in quasi-DAs approaches infinity, the expansion  in terms of $\Lambda_{\rm QCD}/P^z$ yields the LCDA convoluting with a hard kernel at leading power. In these works, a hybrid renormalization scheme, which is composed of self-renormalization scheme and ratio scheme, is adopted to renormalize the spatial correlators~\cite{Han:2023xbl}.
With the self-renormalization,  all divergences except the $\ln (\mu^2z_i^2)$ terms at short distances are removed  with  no additional non-perturbative contributions.
By selecting appropriate matrix elements and taking a ratio with respect to them at short distances, the residual divergences can be eliminated successfully.
The corresponding   hard kernel has also been obtained at the final step. However only one distribution amplitude of the $\Lambda$ baryon  has been studied as an example.

In this paper, we will apply this procedure to all leading-twist LCDAs of lowest-lying octet and decuplet baryons. At leading-twist accuracy, there are three distribution amplitudes for the octet baryons and four for the decuplet baryons. We will provide appropriate  definitions for LCDAs and corresponding quasi-DAs that are suitable for our calculations. As we will point out, the renormalization will only depend on the non-local operators inside the spatial correlators, some of which are the same for octet and decuplet baryons. 
This fact greatly simplifies the investigation.  
After  the renormalization of spatial correlators in the hybrid renormalization scheme, a complete result for the hard kernels will  be obtained at next-to-leading order in $\alpha_s$ which will provide valuable insights for comprehensive  Lattice QCD studies in future.   

The rest of this paper is arranged as follows.
In Sec.~\ref{sec:LCDA}, an overview of the LCDAs of the lowest-lying octet and decuplet baryons is provided. This includes their definitions, symmetry properties, and the expansion in terms of Gegenbauer moments with the aid of conformal spin symmetry.
Sec.~\ref{sec:quasiDA} is dedicated to discussing the corresponding quasi-DAs. Additionally, one-loop results of spatial correlators are presented.
In Sec.~\ref{sec:hybrid}, the spatial correlators are renormalized using the hybrid renormalization scheme, and the corresponding  hard kernels are presented. 
The final section provides a summary. The flavor SU(3) and spin SU(2) wave functions are shown in Appendix~\ref{sec:wf}. Details of the calculation of hard kernels are contained in Appendix~\ref{sec:appe kernel}.

\section{Light baryon lightcone distribution amplitudes} 
\label{sec:LCDA}

In this section, we provide an overview of LCDAs for light baryons. We begin by defining leading-twist LCDAs and presenting specific Dirac matrices to separate them. Then, we explore the symmetry properties of LCDAs and demonstrate how their asymptotic behavior can be understood through conformal spin symmetry.

\subsection{Definitions of LCDAs}

The lightcone distribution amplitude is defined as the hadron-to-vacuum matrix element of non-local operators
\begin{eqnarray}
\varepsilon^{i j k} \times \langle 0 | f_\alpha^{i^{\prime}}\left(z_1 \right)U_{i^{\prime} i}\left(z_1 , z_0 \right)  g_\beta^{j^{\prime}}\left(z_2 \right)
U_{j^{\prime} j}\left(z_2 , z_0 \right) h_\gamma^{k^{\prime}}\left(z_3 \right) U_{k^{\prime} k}\left(z_3 , z_0 \right)
| B(P_B,\lambda)\rangle, \label{eq:lcda_generic}
\end{eqnarray}
where $|B(P_B,\lambda)\rangle$ represents a baryon state with momentum $P_B$, satisfying $P_B^2=M_B^2$, and helicity $\lambda$. 
The indices $\alpha$, $\beta$, and $\gamma$ correspond to Dirac indices. 
The indices $i^{(\prime)}$, $j^{(\prime)}$, and $k^{(\prime)}$ denote color charges, and $f$, $g$, $h$ represent quark fields of each baryon, which are illustrated in TABLE.~\ref{tab:flavor_structure}.

\begin{widetext}
\begin{table}[http]
\centering
\caption{The corresponding valence quark $f,g,h$ for  each light baryon}
\label{tab:flavor_structure}
\begin{tabular}{c |c| c| c| c| c| c| c| c|c|c|c}
\hline 
 Octet & $n$ & $p$ & $ \Sigma ^{-}$ & $\Sigma^{0}$ & $ \Sigma ^{+}$ & $ \Xi ^{-}$ & $ \Xi ^{0}$ &  &  & & $\Lambda$ \\
\hline 
 Decuplet & $\displaystyle \Delta ^{0}$ & $\displaystyle \Delta ^{+}$ & $\displaystyle \Sigma ^{*-}$ & $\displaystyle \Sigma ^{*0}$ & $\displaystyle \Sigma ^{*+}$ & $\displaystyle \Xi ^{*-}$ & $\displaystyle \Xi ^{*0}$ & $\displaystyle \Delta ^{++}$ & $\displaystyle \Delta ^{-}$ & $\displaystyle \Omega ^{-}$ \\
\hline 
 f g h & d d u & u u d & d d s & $\frac{1}{\sqrt 2}$(u d s+d u s) & u u s & s s d & s s u & u u u & d d d & s s s & u d s \\
 \hline
\end{tabular}
\end{table}
\end{widetext}

The coordinates $z_i$ are set on the light cone with $z_i^\mu=z_i n^\mu$, where $z_i^2=0$. In this context, two lightcone unit vectors are defined as $n^\mu=(1,0,0,-1)/\sqrt{2}$ and $\bar n^\mu=(1,0,0,1)/\sqrt{2}$. The momentum of the baryon is directed along the $\bar n$ direction, expressed as $P^\mu=P^{+} \bar n^\mu = (P^z,0,0,P^z)$.
The Wilson lines $U(x,y)$ are introduced to maintain  gauge invariance
\begin{equation}
U(x, y)=\mathcal{P} \exp \left[i g \int_0^1 \mathrm{~d} t(x-y)_\mu A^\mu(t x+(1-t) y)\right].
\end{equation}
In Eq.~\eqref{eq:lcda_generic}, the starting position $z_0$ in Wilson lines can be arbitrarily chosen along the direction of $n^\mu$ due to the gauge invariance.  For the sake of simplicity, we set $z_0=0$ in the subsequent discussion. Additionally, unless necessary, we will omit the explicit display of Wilson lines and color indices.

At  leading twist, the LCDAs for an octet baryon can be decomposed into three terms as~\cite{Chernyak:1984bm}
\begin{align}\label{def-oct}
&  \left\langle 0\left| f_\alpha\left(z_1 n\right) g_\beta\left(z_2 n\right) h_\gamma\left(z_3 n\right)\right| B\left(P_B, \lambda\right)\right\rangle\notag \\&
=\frac{1}{4}f_V\left[\left(P\!\!\!\!/_B C\right)_{\alpha \beta}\left(\gamma_5 u_B\right)_\gamma V^B\left(z_i n \cdot P_B\right)\right. 
\left.+\left( P\!\!\!\!/_B \gamma_5 C\right)_{\alpha \beta} (u_B)_\gamma A^B\left(z_i n \cdot P_B\right) \right] \nonumber
\\&
+\frac{1}{4}f_T\left(i \sigma_{\mu \nu} P_B^\nu C\right)_{\alpha \beta}\left(\gamma^\mu \gamma_5 u_B\right)_\gamma T^B\left(z_i n \cdot P_B\right),
\end{align}
where $C\equiv i \gamma^2\gamma^0$ is the charge conjugation matrix,
$u_{B}$ stands for the  spinor for the  baryon and $\sigma_{\mu \nu}=\frac{i}{2}\left[\gamma_\mu, \gamma_\nu\right]$. $f_{V/A/T}$ is the corresponding decay constant for each LCDA.  
For proton and neutron, $f_T=f_V$ due to the isospin symmetry.

Similarly, the LCDA for a decuplet baryon can be decomposed into four terms as~\cite{Farrar:1988vz}
\begin{align}
&  \langle 0|  f_\alpha \left(z_1 n\right) g_\beta \left(z_2 n\right) h_\gamma\left(z_3 n\right)\left|B\left(P_{B}, \lambda\right)\right\rangle 
\notag\\&
=\frac{1}{4}\lambda_{V}\left[\left(\gamma_\mu C\right)_{\alpha \beta} \Delta_\gamma^\mu V^B\left(z_i n \cdot P_{B}\right)
+\left(\gamma_\mu \gamma_5 C\right)_{\alpha \beta}\left(\gamma_5 \Delta^\mu\right)_\gamma A^B\left(z_i n \cdot P_{B}\right)\right] 
\notag\\&-\frac{1}{8}\lambda_T\left(i \sigma_{\mu \nu} C\right)_{\alpha \beta}\left(\gamma^\mu \Delta^\nu\right)_\gamma T^{B}\left(z_i n \cdot P_{B}\right)-\frac{1}{4}\lambda_{\varphi}\left[\left(i \sigma_{\mu \nu} C\right)_{\alpha \beta}\left(P_B^\mu \Delta^\nu-\frac{1}{2}M_B\gamma^\mu\Delta^\nu\right)_\gamma \varphi^B\left(z_i n \cdot P_{B}\right)\right],
\end{align}
where $\Delta^\mu_\gamma$ is the abbreviation for  spin-$\frac{3}{2}$ vector $\Delta^\mu_\gamma(p,\lambda)$, which satisfies: 
$ (p\!\!/-M_B) \Delta^\mu_\gamma(p,\lambda)=0, \quad \bar{\Delta}_\mu^\gamma(p,\lambda)\Delta^\mu_\gamma(p,\lambda)=-2M_B,\quad \gamma_\mu \Delta^\mu_\gamma(p,\lambda)=0,\quad p_\mu \Delta^\mu_\gamma(p,\lambda)=0$. 
The spin-$\frac{3}{2}$ vector can be expressed with spinors and polarized vectors
\begin{equation}
\begin{aligned}
\Delta_{\mu\gamma}(p, \lambda=1 / 2)=\sqrt{\frac{2}{3}} e_\mu^0(p) u_\gamma^{\uparrow}(p)+\sqrt{\frac{1}{3}} e_\mu^{+1}(p) u_\gamma^{\downarrow}(p), \;\; \Delta_\gamma^\mu(p, \lambda=3 / 2)=e_\mu^{+1}(p) u_\gamma^{\uparrow}(p).
\end{aligned}
\end{equation}
Up to leading twist, the light baryon mass can be neglected $M_B\simeq 0$ and the polarized vectors are $
e^0_\mu\simeq\bar{n};e^+_\mu=(0,1,-i,0);e^-_{\mu}=(0,1,i,0)$. $\lambda_{V/A/T/\varphi}$ is the corresponding decay constant for each LCDA.
In the decomposition of decuplet baryons, $V,A,T$ correspond to helicity-$1/2$ state while $\varphi$ corresponds to helicity-$3/2$ state.

For pratical use we can choose appropriate Dirac matrices to project out $V$, $A$, $T$ and $\varphi$, respectively. 
For the octet baryons, the $V$, $A$ and  $T$ can be given through 
\begin{align}\label{lc-oct-de}
&\left\langle 0\left|f^T\left(z_1 n\right) 
(C n\!\!\!/)
g\left(z_2 n\right) h\left(z_3 n\right)
\right| B\right\rangle
=-f_{V} V^B(z_i n \cdot P_B)P_B^+\gamma_5 u_{B},
\notag\\&
\left\langle 0\left|f^T\left(z_1 n\right) 
(C \gamma_5  n\!\!\!/)
g\left(z_2 n\right)  h\left(z_3 n\right)
\right| B\right\rangle
=f_{V} A^B(z_i n \cdot P_B) P_B^+u_{B},
\notag\\&
\left\langle 0\left|f^T\left(z_1 n\right) 
(i C \sigma_{\mu\nu} n^\nu )
g\left(z_2 n\right)  \gamma^\mu h\left(z_3 n\right)
\right| B\right\rangle
= 2f_{T} T^B(z_i n \cdot P_B)P_B^+\gamma_5 u_{B},
\end{align}
with the normalization
\begin{equation}
\begin{aligned}
    &V^{B\neq \Lambda}(0,0,0)=T^{B\neq \Lambda}(0,0,0)=1,\quad A^{B\neq \Lambda}(0,0,0)=0,
      \quad
    V^{\Lambda}(0,0,0)=T^{\Lambda}(0,0,0)=1,\quad A^{ \Lambda}(0,0,0)=0.
    \end{aligned}
\end{equation}
For the decuplet baryons, $V,A,T$ and $\varphi$ can be given by
\begin{align}\label{lc-dec-de}
& \left\langle 0\left| f^T\left(z_1 n\right)(C  n\!\!\!/) g\left(z_2 n\right) h\left(z_3 n\right)\right| B\left(P_B, \frac{1}{2}\right)\right\rangle
=- \lambda_{V} V^B\left(z_i n \cdot P_B\right)\gamma_5 (n \cdot \Delta),
\notag\\
& \left\langle 0\left| f^T\left(z_1 n\right)(C \gamma_5 n\!\!\!/) g\left(z_2 n\right) h\left(z_3 n\right)\right| B\left(P_B, \frac{1}{2}\right)\right\rangle =\lambda_{V} A^B\left(z_i n \cdot P_B\right)(n \cdot \Delta),
\notag\\
& \left\langle 0\left| f^T\left(z_1 n\right)( iC \sigma_{\mu\nu} n^\nu ) g\left(z_2 n\right) \gamma^\mu h\left(z_3 n\right)\right| B\left(P_B, \frac{1}{2}\right)\right\rangle =- \lambda_{T} T^B\left(z_i n \cdot P_B\right) (n \cdot \Delta),
\notag\\
 &\left\langle 0\left|f^T\left(z_1 n\right)\left(iC \sigma^{\mu \nu} n_\mu\right)  g\left(z_2 n\right) h\left(z_3 n\right)\right| B\left(P_B , \frac{3}{2}\right)\right\rangle 
=-\lambda_{\varphi}\varphi^B\left(z_i n \cdot P_B\right) P_B^+\Delta^\nu,
\end{align}
with the normalization
\begin{equation}
\begin{aligned}
V^{\Lambda}(0,0,0)=T^{\Lambda}(0,0,0)=\varphi^B(0,0,0)=1,\quad A^{ \Lambda}(0,0,0)=0.
\end{aligned}
\end{equation}

As long as $V,A,T$ and $\varphi$ does not vanish at $z_1=z_2=z_3=0$, decay constants can be given through
\begin{align}
&\left\langle 0\left|f^T\left(0\right) 
(C n\!\!\!/)
g\left(0\right) h\left(0\right)
\right| B\neq \Lambda(P_B,\frac{1}{2})\right\rangle
=-f_{V} P_B^+\gamma_5 u_{B},
\notag\\&
\left\langle 0\left|f^T\left(0\right) 
(C \gamma_5  n\!\!\!/)
g\left(0\right)  h\left(0\right)
\right| \Lambda(P_B,\frac{1}{2})\right\rangle
=f_{V}  P_B^+u_{B},
\notag\\&
\left\langle 0\left|f^T\left(0\right) 
(i C \sigma_{\mu\nu} n^\nu )
g\left(0\right)  \gamma^\mu h\left(0\right)
\right| B\neq\Lambda(P_B,\frac{1}{2})\right\rangle
= 2f_{T}  P_B^+\gamma_5 u_{B},
\notag\\
& \left\langle 0\left| f^T\left(0\right)(C  n\!\!\!/) g\left(0\right) h\left(0\right)\right| B\left(P_B, \frac{1}{2}\right)\right\rangle 
=- \lambda_{V}  \gamma_5 (n \cdot \Delta),
\notag\\
& \left\langle 0\left| f^T\left(0\right)(C \gamma_5 n\!\!\!/) g\left(0\right) h\left(0\right)\right| B\left(P_B, \frac{1}{2}\right)\right\rangle =\lambda_{V} (n \cdot \Delta),
\notag\\
& \left\langle 0\left| f^T\left(0\right)( iC \sigma_{\mu\nu} n^\nu ) g\left(0\right) \gamma^\mu h\left(0\right)\right| B\left(P_B, \frac{1}{2}\right)\right\rangle =- \lambda_{T} (n \cdot \Delta),
\notag\\
 &\left\langle 0\left|f^T\left(0\right)\left(iC \sigma^{\mu \nu} n_\mu\right)  g\left(0\right) h\left(0\right)\right| B\left(P_B , \frac{3}{2}\right)\right\rangle 
=-\lambda_{\varphi} P_B^+\Delta^\nu. 
\end{align}

For later convenience, the LCDAs in momentum space are also defined
\begin{equation}\label{eq:LCDA}
\begin{aligned}
& \Phi^B\left(x_1, x_2,x_3,\mu \right)=
\int_{-\infty}^{+\infty} \frac{n\cdot P d \,  z_1}{2 \pi}  \frac{n\cdot P d \, z_2}{2 \pi}
\times  e^{i x_1 n\cdot P z_1+i x_2 n\cdot P z_2} 
\times \Phi^B_R\left(z_1 n\cdot P, z_2 n\cdot P,0,\mu\right),
\end{aligned}
\end{equation}
where $\Phi_R^B$ stands for the renormalized spatial LCDAs, $V,A,T$ and $\varphi$, and $\mu$ is the renormalization scale. The symbols $x_1,x_2,x_3(=1-x_1-x_2)$ label the longitudinal momentum fractions carried by three valence quarks, satisfying $0 \leq x_i \leq 1$ and $\sum_ix_i=1$. In the following discussion, the subscript ``R" will be omitted, and the parameter $\mu$ will denote renormalized quantities accordingly.

Additionally, an alternative basis can be defined for the octet baryon~\cite{Wein:2015oqa,Bali:2015ykx}: 
\begin{align}\label{alt-bas}
\Phi_{ \pm}^{B \neq \Lambda}\left(x_1,x_2,x_3\right) & =\frac{1}{2}\left([V-A]^B\left(x_1,x_2,x_3\right) \pm[V-A]^B\left(x_3,x_2,x_1\right)\right), 
\notag\\
\Pi^{B \neq \Lambda}\left(x_1,x_2,x_3\right) & =T^B\left(x_1,x_3,x_2\right), \notag\\
\Phi_{+}^{\Lambda}\left(x_1,x_2,x_3\right) & =\sqrt{\frac{1}{6}}\left([V-A]^{\Lambda}\left(x_1,x_2,x_3\right)+[V-A]^{\Lambda}\left(x_3,x_2,x_1\right)\right), \notag\\
\Phi_{-}^{\Lambda}\left(x_1,x_2,x_3\right) & =-\sqrt{\frac{3}{2}}\left([V-A]^{\Lambda}\left(x_1,x_2,x_3\right)-[V-A]^{\Lambda}\left(x_3,x_2,x_1\right)\right), \notag\\
\Pi^{\Lambda}\left(x_1,x_2,x_3\right) & =\sqrt{6} T^{\Lambda}\left(x_1,x_3,x_2\right).
\end{align}
This basis  is intricately connected to the light-front wave functions defined in the next section, which is advantageous to explore $\rm{SU(3)}$ flavor symmetry and the potential violation.

\subsection{Symmetry properties of LCDAs}

The LCDAs exhibit specific symmetry properties based on the flavor structures and  Lorentz structures in their operators~\cite{Braun:1999te}. Further exploration involving isospin symmetry and the complete $\rm{SU(3)}$ flavor symmetry reveals additional symmetry properties inherent in the $V, A, T$ and $\varphi$. Through lattice calculations, these symmetry properties can be scrutinized, providing a means to evaluate the magnitudes of isospin and $\rm{SU(3)}$ flavor symmetry breaking effects. In the following, the symmetry properties of octet baryons will be presented based on identical fields, isospin symmetry, and the limit of $\rm{SU(3)}$ flavor symmetry, while the symmetry properties of decuplet baryons will be categorized according to the relations they fulfill.

\subsubsection{Octet baryons}
\begin{itemize}
    \item \textbf{Identical  quark fields}
\end{itemize}
Given that all octet baryons, except the $\Lambda$ and $\Sigma^0$, $V,A,T$ exhibit specific symmetries under the interchange of $x_1$ and $x_2$:
\begin{align}
& V^B\left(x_1, x_2, x_3\right)=V^B\left(x_2, x_1, x_3\right),\quad
A^B\left(x_1, x_2, x_3\right)=-A^B\left(x_2, x_1, x_3\right), \quad
T^B\left(x_1, x_2, x_3\right)=T^B\left(x_2, x_1, x_3\right) .
\end{align}
Here $B$ stands for an octet baryon expect $\Lambda$ and $\Sigma^0$.
\begin{itemize}
    \item \textbf{Isospin symmetry}
\end{itemize}

For proton and neutron, further imposition of isospin symmetry leads to the constraint
\begin{equation}\label{2TVA}
2 T^{n,p}\left(x_1, x_2, x_3\right)=[V-A]^{n,p}\left(x_1, x_3, x_2\right)+[V-A]^{n,p}\left(x_2, x_3, x_1\right).
\end{equation}

For $\Lambda$ and $\Sigma^0$,  one has the corresponding constraint
\begin{align}\label{eq-iso}
& V^\Lambda \left(x_1, x_2, x_3\right)=-V^\Lambda \left(x_2, x_1, x_3\right),\quad
A^\Lambda \left(x_1, x_2, x_3\right)=A^\Lambda \left(x_2, x_1, x_3\right), \quad
T^\Lambda \left(x_1, x_2, x_3\right)=-T^\Lambda \left(x_2, x_1, x_3\right),
\notag\\
& V^ {\Sigma^0} \left(x_1, x_2, x_3\right)=V^ {\Sigma^0} \left(x_2, x_1, x_3\right),\quad
A^ {\Sigma^0} \left(x_1, x_2, x_3\right)=-A^ {\Sigma^0} \left(x_2, x_1, x_3\right), \quad
T^ {\Sigma^0} \left(x_1, x_2, x_3\right)=T^ {\Sigma^0} \left(x_2, x_1, x_3\right).
\end{align}

\begin{itemize}
    \item \textbf{$\rm{SU(3)}$ flavor symmetry}
\end{itemize}

In the limit of $\rm{SU(3)}$ flavor symmetry, Eq.~(\ref{2TVA})  holds true for all baryons
\begin{align}
    &2 T^{\Lambda}\left(x_1, x_2, x_3\right)=-[V-A]^{\Lambda}\left(x_1, x_3, x_2\right)+[V-A]^{\Lambda}\left(x_2, x_3, x_1\right),
    \notag\\
    &2 T^{ B\neq \Lambda}\left(x_1, x_2, x_3\right)=[V-A]^{ {B\neq \Lambda}}\left(x_1, x_3, x_2\right)+[V-A]^{ B\neq \Lambda}\left(x_2, x_3, x_1\right).
\end{align}

In the alternative basis defined in Eq.~(\ref{alt-bas}), the $SU(3)$ flavor symmetry gives
\begin{equation}
\begin{aligned}
& \Phi_{+}\equiv \Phi_{+}^{n,p }=\Phi_{+}^{\Sigma^{+/0/-}  }=\Phi_{+}^{\Xi^{0/-}  }=\Phi_{+}^{\Lambda  }=\Pi^{n,p  }=\Pi^{\Sigma^{+/0/-}  }=\Pi^{\Xi^{0/-}  }, \\
& \Phi_{-}^{ } \equiv \Phi_{-}^{n,p  }=\Phi_{-}^{\Sigma^{+/0/-}  }=\Phi_{-}^{\Xi^{0/-}  }=\Phi_{-}^{\Lambda  }=\Pi^{\Lambda  } .
\end{aligned}
\end{equation}
where $B$ stands for all octet baryons.

These listed symmetry properties can also be derived using light-front wave functions. If one only considers  the S-wave contribution where the helicity of baryons is totally provided by three valence quarks, by combining the color, momentum, flavor and spin wave functions, one can construct the light-front wave functions~\cite{Wein:2015oqa}:
\begin{align}
\left|B \neq (\Lambda)^{\uparrow}\right\rangle
=\int \frac{[d x]}{8 \sqrt{3 x_1 x_2 x_3}}|\uparrow \uparrow \downarrow\rangle \otimes\{&-\sqrt{3} \Phi_{+}^B\left(x_1, x_3, x_2\right)(|{MS}, B\rangle-\sqrt{2}|{S}, B\rangle) / 3 
\notag\\
& -\sqrt{3} \Pi^B\left(x_1, x_3, x_2\right)(2|{MS}, B\rangle+\sqrt{2}|{S}, B\rangle) / 3 
\notag\\
&\left.+\Phi_{-}^B\left(x_1, x_3, x_2\right)|{MA}, B\rangle\right\},\\
\left|\Lambda^{\uparrow}\right\rangle
=\int \frac{[d x]}{8 \sqrt{3 x_1 x_2 x_3}}|\uparrow \uparrow \downarrow\rangle \otimes\{ & -\sqrt{3} \Phi_{+}^{\Lambda}\left(x_1, x_3, x_2\right)|{MS}, \Lambda\rangle 
\notag\\
+ & \Pi^{\Lambda}\left(x_1, x_3, x_2\right)(2|{MA}, \Lambda\rangle+\sqrt{2}|{A}, \Lambda\rangle) / 3 
\notag\\
+ & \left.\Phi_{-}^{\Lambda}\left(x_1, x_3, x_2\right)(|{MA}, \Lambda\rangle-\sqrt{2}|{A}, \Lambda\rangle) / 3\right\}.
\end{align}
$[dx]$ stands for $dx_1dx_2dx_3$ and the arrows denote the helicity of the quarks. The notation $MS/MA$ denotes the octet flavor wave function, which is even/odd with respect to the first two quarks. The notation $S/A$ refers to the symmetric/antisymmetric flavor wave functions, which arise only when the $\rm{SU(3)}$ flavor symmetry is broken. For detailed definitions of the spin and flavor wave functions, please refer to Appendix~\ref{sec:wf}.

To compare the light-front wave function with LCDAs, we can alternatively define LCDAs in Eq.~(\ref{def-oct}) as  the states
\begin{align}
|(B=\Lambda/\Sigma^0)^
{\uparrow}\rangle=\int \frac{[d x]}{4 \sqrt{6 x_1 x_2 x_3}}|u d s\rangle \otimes\{ &f_V[V+A]^{\Lambda}(x_1, x_2, x_3)|\downarrow \uparrow \uparrow\rangle+f_V[V-A]^{\Lambda}(x_1, x_2, x_3)|\uparrow \downarrow \uparrow\rangle \}   \notag\\
-&2f_T T^{\Lambda}(x_1, x_2, x_3)|\uparrow \uparrow \downarrow\rangle\}, \\
|(B\neq\Lambda/\Sigma^0)^{\uparrow}\rangle
=\int \frac{[d x]}{8 \sqrt{6 x_1 x_2 x_3}}|f g h\rangle \otimes\{&f_V[V+A]^B(x_1, x_2, x_3)|\downarrow \uparrow \uparrow\rangle+f_V[V-A]^B\left(x_1, x_2, x_3\right)|\uparrow \downarrow \uparrow\rangle\notag\\
-&2f_T T^B(x_1, x_2, x_3)|\uparrow \uparrow \downarrow\rangle\} .
\end{align}
The overall factor of $\Lambda/\Sigma^0$ is twice that of other octet baryons to compensate for two different ways of contraction arising from two identical quark fields in $B\neq \Lambda/\Sigma^0$.

\subsubsection{Decuplet baryons}

The analysis for the  decuplet baryons is  similar to that of the octet baryons. For simplicity, we divide this subsection according to the relations.

\begin{itemize}
    \item \textbf{$x_1\leftrightarrow x_2$ exchange in V,A,T, and $\varphi$}
\begin{align}
& V^B\left(x_1, x_2, x_3\right)=V^B\left(x_2, x_1, x_3\right),\;\;\;
A^B\left(x_1, x_2, x_3\right)=-A^B\left(x_2, x_1, x_3\right), \notag\\&
T^B\left(x_1, x_2, x_3\right)=T^B\left(x_2, x_1, x_3\right) ,\;\;\;
\varphi^B\left(x_1, x_2, x_3\right)=\varphi^B\left(x_2, x_1, x_3\right). \label{eq:decuplet_LCDA_relation}
\end{align}
    \begin{enumerate}
        \item Due to two identical quark fields, Eq.~(\ref{eq:decuplet_LCDA_relation}) holds for  decuplet baryons other than $\Sigma^{*0}$ .
        \item When isospin symmetry is assumed, it also holds for $\Sigma^{*0}$.
    \end{enumerate}

    \item \textbf{Further relation}
 \begin{eqnarray}\label{further-1}
    T^{B}\left(x_1, x_2, x_3\right)=[V-A]^{B}\left(x_2, x_3, x_1\right),\\
\label{further-2}
\varphi^B(x_1,x_2,x_3)=\varphi^B(x_2,x_3,x_1)=\varphi^B(x_3,x_1,x_2),
\end{eqnarray}

 which indicates that $\varphi^B$ is totally symmetric.
    \begin{enumerate}
        \item Due to three identical quark fields, Eqs.~(\ref{further-1}, \ref{further-2}) hold for $\Delta^{++}$,$\Delta^+$, and $\Omega^-$ .
        \item When isospin symmetry is assumed, Eqs.~(\ref{further-1}, \ref{further-2}) also hold for $\Delta^0$ and $\Delta^+$.
        \item In the limit of $\rm{SU(3)}$ flavor symmetry, Eqs.~(\ref{further-1}, \ref{further-2}) hold for all decuplet baryons.
    \end{enumerate}
\end{itemize}

\subsection{Conformal spin expansion of LCDAs}

Despite QCD being not a conformal field theory, conformal spin symmetry still influences QCD~\cite{Braun:2003rp}. This influence stems from a significant observation: up to one loop accuracy, the renormalization of operators is governed by tree-level counterterms that inherit conformal symmetry from the classical Lagrangian. In the framework of conformal field theory, we employ primary fields labeled by conformal spins as the building blocks of composite operators. Notably, the renormalization group evolution (RG), which can be viewed as an integral operator, commutes with the $\mathfrak{sl}(2,\mathbb{R})$ collinear subalgebra. Consequently, operators with distinct conformal spins do not undergo mixing with each other under RG. Therefore, to investigate the scale dependence of LCDAs, it is advantageous to express the LCDAs as expansions in terms of operators with definite conformal spins
\begin{equation}\label{cm}
\Phi^B_R(x_1,x_2,x_3,\mu)=120x_1x_2x_3\sum_{N=0}^{\infty}\sum_{q=0}^N\phi_{N,q}(\mu_0)P_{N,q}(x_1,x_2,x_3)\left(\frac{\alpha_s(\mu)}{\alpha_s(\mu_0)}\right)^{\gamma_{N,q}/\beta_0},
\end{equation}
where $\Phi^B_R$ is the renormalized $V,A,T$, or $\varphi$, and $\phi_{N,q}(\mu_0)$ are called the conformal moments corresponding to the operators with conformal spins $\frac{3}{2}+N$~\cite{Anikin:2013yoa,Braun:2008ia}. Eq.~(\ref{cm}) illustrates how LCDAs evolve from scale $\mu_0$ to scale $\mu$. $P_{N,q}(x_1,x_2,x_3)$ are homogeneous functions which define the corresponding conformal moments. They are orthogonal with respect to
\begin{equation}\label{orthogonal}
\int_0^1[dx]x_1x_2x_3P_{N,q}(x_1,x_2,x_3)P_{N',q'}(x_1,x_2,x_3)= c_{N,q}\delta_{qq}\delta_{NN'}.
\end{equation}
Eq.~(\ref{orthogonal})  does not uniquely determine the polynomials, as they depend on the helicity of the baryons. For helicity-3/2 states, $P_{N,q}(x_1,x_2,x_3)$ and $c_{N,q}$ are exactly solvable. For helicity-1/2 states, however, they can be only solved numerically. Fortunately, the first few polynomials of helicity-1/2 and helicity-3/2 states are numerically similar. The detailed determination of 
 $P_{N,q}(x_1,x_2,x_3)$ can be found in~\cite{Braun:2003rp}.

In QCD, the anomalous dimension $\gamma_{N,q}$ in Eq.~(\ref{cm}) increases with the conformal spin. So the behaviour of LCDAs at large $\mu$ is governed by the first few terms. Since both LCDAs and $P_{N,q}$ have definite symmetry in $x_1$ and $x_2$, some terms in the expansion automatically vanish~\cite{Braun:2008ia}. 

\begin{itemize}
    \item 
    For octet and decuplet baryons $B \neq \Lambda$, the expansions are 
\begin{align}
V^B\left(x_1,x_2,x_3, \mu\right) & =120 x_1 x_2 x_3\left[\phi_{0,0}^{v}(\mu)+\phi_{1,1}^{v}(\mu)\left(1-3 x_3\right)+...\right], \notag\\
A^B\left(x_1,x_2,x_3, \mu\right) & =120 x_1 x_2 x_3[\phi_{1,0}^a(\mu)\left(x_2-x_1\right) +...], \notag\\
T^B\left(x_1,x_2,x_3, \mu\right) & =120 x_1 x_2 x_3\left[\phi_{0,0}^t(\mu)+\phi_{1,0}^t(\mu)\left(1-3 x_3\right)+...\right],
\end{align}
where $\phi_{N,q}^{v/a/t}(\mu)$ is the moment for $V^B/A^B/T^B$ and the ellipsis stands for terms suppressed by higher conformal spins.
\item 
For neutron and proton, due to isospin symmetry,  $T^{n/p}(x_1,x_2,x_3,\mu)$ can be expressed as 
\begin{align}
T^{n/p}\left(x_1,x_2,x_3, \mu\right) & =120 x_1 x_2 x_3\left[\phi_{0,0}^v(\mu)+\frac{1}{2}\left(\phi_{1,0}^{a}-\phi_{11}^{v}\right)(\mu)\left(1-3 x_3\right)+...\right] .
\end{align}

\item 

The expansions of LCDAs for $\Lambda$ are
\begin{align}
V^\Lambda\left(x_1,x_2,x_3, \mu\right) & =120 x_1 x_2 x_3
[\phi_{1,0}^v(\mu)\left(x_2-x_1\right) +...],
\notag\\
A^\Lambda\left(x_1,x_2,x_3, \mu\right) & =120 x_1 x_2 x_3
[\phi_{0,0}(\mu)^a+\phi_{1,1}^a(\mu)(1-3 x_3) +...],
\textbf{}\notag\\
T^\Lambda\left(x_1,x_2,x_3, \mu\right) & =120 x_1 x_2 x_3
 [\phi_{1,0}^t(\mu)\left(x_2-x_1\right)+...] .
\end{align}

\item 
For the $\varphi^{B}$ of decuplet baryons, the expansions are
\begin{equation}
\begin{aligned}
\varphi^B\left(x_1,x_2,x_3, \mu\right) & =120 x_1 x_2 x_3\left[\phi_{0,0}(\mu)+\phi_{1,0}(\mu)\left(1-3 x_3\right)+...\right] .
\end{aligned}
\end{equation}

\item 
For the alternative basis defined in Eq.~(\ref{alt-bas}), the corresponding expansions are
\begin{align}
\Phi_{+}^B(x_1,x_2,x_3,\mu) & =120 x_1 x_2 x_3\left(\varphi_{0,0}^B (\mu)\mathcal{P}_{00}(x_1,x_2,x_3)+\varphi_{1,1}^B (\mu)\mathcal{P}_{11}(x_1,x_2,x_3)+\ldots\right), \notag\\
\Phi_{-}^B(x_1,x_2,x_3,\mu) & =120 x_1 x_2 x_3\left(\varphi_{1,0}^B (\mu)\mathcal{P}_{10}(x_1,x_2,x_3)+\ldots\right), \notag\\
\Pi^{B \neq \Lambda}(x_1,x_2,x_3,\mu) & =120 x_1 x_2 x_3\left(\pi_{0,0}^B (\mu)\mathcal{P}_{00}(x_1,x_2,x_3)+\pi_{1,1}^B (\mu)\mathcal{P}_{11}(x_1,x_2,x_3)+\ldots\right), \notag\\
\Pi^{\Lambda} (x_1,x_2,x_3,\mu)& =120 x_1 x_2 x_3\left(\pi_{1,0}^A (\mu)\mathcal{P}_{10}(x_1,x_2,x_3)+\ldots\right),
\end{align}
where $\varphi^B_{N,q}(\mu)$ and $\pi^B_{N,q}(\mu)$ are the conformal moments defined in the alternative basis. The explicit forms for $\mathcal{P}_{00,10,11}$ are given as:
\begin{equation}
    \mathcal{P}_{00}=1,\quad
    \mathcal{P}_{10}=21(x_1-x_3),\quad \mathcal{P}_{11}=7(x_1-2x_2+x_3).
\end{equation}

\end{itemize}

\section{SPATIAL CORRELATOR and Quasi Distribution Amplitudes}
\label{sec:quasiDA}
In this section, we start the process of the extraction of the LCDAs from the spatial correlators, utilizing LaMET. 
To lay down the foundation, a brief review of LaMET is presented. 
Subsequently, explicit definitions of the quasi-DAs are provided. 
The  renormalization and matching procedures will be addressed in the next section.

\subsection{ A brief review of LaMET formalism}
 
Within LaMET, the determination of the LCDA begins with the construction of a spatial correlator. 
The Fourier transformation of this correlator is referred to as a quasi-DA.
Sharing the same infrared (IR) structure as the corresponding LCDAs, quasi-DAs can be effectively simulated with Lattice QCD. 
In the large momentum limit, their distinctions lie in the ultraviolet (UV) structure which is perturbatively calculable, and the collinear part of a quasi-DA can be identifies as a LCDA. This leads to the factorization at the leading order:
\begin{align}\label{matching}
\tilde{\Phi}^B\left(x_1, x_2, P_B^z,\mu\right)= \int d y_1 d y_2 \mathcal{C}\left(x_1, x_2,y_1, y_2,P_B^z,\mu\right) \Phi^B\left(y_1, y_2,\mu\right) +\mathcal{O}\left(\frac{\Lambda^2_{\rm QCD}}{(x_1 P_B^z)^2}, \frac{\Lambda^2_{\rm QCD}}{(x_2 P_B^z)^2}, \frac{\Lambda^2_{\rm QCD}}{(\left(1-x_1-x_2\right) P_B^z)^2}\right).
\end{align}
Here, $P^z_B$ is the momentum of the hadron along the $z$ direction, $\tilde{\Phi}$ stands for $\tilde{V}$, $\tilde{A}$, $\tilde{T}$ and $\tilde{\varphi}$, the corresponding quasi-DAs of $V, A, T$ and $\varphi$, and the $x_3(y_3)$ argument is omitted for simplicity. $\mathcal{C}(x_1,x_2,y_1,y_2,P_B^z,\mu)$, referred to as the hard kernel, compensates for the UV difference between   LCDAs and quasi-DAs. The spirit of Eq.~(\ref{matching}) lies in substituting an LCDA for a quasi-DA, which is computable using Lattice QCD, along with a hard kernel that is perturbative calculable. 
The hard kernel for one distribution amplitude $A$ of the $\Lambda$ baryon, has been calculated up to one-loop accuracy in the $\overline {\rm MS}$ scheme~\cite{Deng:2023csv}. The $\mu$ in the quasi-DA originates from the renormalization of the logarithmic divergences, whereas the scale $\mu$ in the LCDA represents the factorization scale utilized to separate the collinear and hard modes. Consequently, the hard kernel $\mathcal{C}$ incorporates both renormalization and factorization scales, which are chosen identical for convenience.

\subsection{Definitions of quasi-DAs}
In this  subsection, we give the definitions of quasi-DAs for octet baryons $B$: 
\begin{align}\label{quasi-def-1}
&\widetilde{M}^B_V(z_1,z_2,z_3,P_B^{z})=\left\langle 0\left|f^T\left(z_1 n_z\right) 
(C \gamma^z)
g\left(z_2 n_z\right)  h\left(z_3 n_z\right)
\right| B(P_B,\lambda=\frac{1}{2})\right\rangle
=- {f_{V}} \tilde V^B(z_1,z_2,z_3,P_B^z) P_B^z\gamma_5 u_{B},
\notag\\&
\widetilde{M}^B_A(z_1,z_2,z_3,P_B^{z})=\left\langle 0\left|f^T\left(z_1 n_z\right) 
(C \gamma_5 \gamma^z)
g\left(z_2 n_z\right)  h\left(z_3 n_z\right)
\right| B(P_B,\lambda=\frac{1}{2}) \right\rangle
= {f_{A}} \tilde A^B(z_1,z_2,z_3,P_B^z) P_B^z u_{B},
\notag\\&
\widetilde{M}^B_T(z_1,z_2,z_3,P_B^{z})=\left\langle 0\left|f^T\left(z_1 n_z\right) 
(  \frac{1}{2} C[\gamma^z,\gamma^\mu])
g\left(z_2 n_z\right)  \gamma_\mu h\left(z_3 n_z\right)
\right| B(P_B,\lambda=\frac{1}{2})\right\rangle
=2  {f_{T}} \tilde T^B(z_1,z_2,z_3,P_B^z) P_B^z \gamma_5 u_{B},
\end{align}
the coordinates are defined as $z_i^\mu=z_i n_z^\mu $, where $n_z^\mu=(0,0,0,1)$. To be complete we also provide the definition of quasi-DAs for decuplet baryons
\begin{equation}
\begin{aligned}\label{quasi-def-2}
& \widetilde{M}^B_V(z_1,z_2,z_3,P_B^{z})=\left\langle 0\left| (f\left(z_1 n_z\right))^T(C \gamma^z) g\left(z_2 n_z\right) h\left(z_3 n_z\right)\right| B(P_B,\lambda=\frac{1}{2})\right\rangle =- \lambda_{V} \tilde V^B\left(z_1,z_2,z_3,P_B^z\right)\gamma_5 (n_z \cdot \Delta),
\\
& \widetilde{M}^B_A(z_1,z_2,z_3,P_B^{z})=\left\langle 0\left| (f\left(z_1 n_z\right))^T(C \gamma_5 \gamma^z) g\left(z_2 n_z\right) h\left(z_3 n_z\right)\right| B(P_B,\lambda=\frac{1}{2})\right\rangle =\lambda_A \tilde A^B\left(z_1,z_2,z_3,P_B^z\right)(n_z \cdot \Delta),
\\
& \widetilde{M}^B_T(z_1,z_2,z_3,P_B^{z})=\left\langle 0\left| (f\left(z_1 n_z\right))^T(\frac{1}{2}C [\gamma^z,\gamma^\mu])  g\left(z_2 n_z\right) \gamma_\mu h\left(z_3 n_z\right)\right| B(P_B,\lambda=\frac{1}{2})\right\rangle =- \lambda_{T} \tilde T^B\left(z_1,z_2,z_3,P_B^z\right) (n_z \cdot \Delta).
\end{aligned}
\end{equation}
The definition of a quasi-DA for helicity $\lambda=3/2$ is
\begin{equation}
\begin{aligned}
&\widetilde{M}^B_\varphi(z_1,z_2,z_3,P_B^{z})=\left\langle 0\left|(f\left(z_1 n_z\right))^T\left(\frac{1}{2}C [\gamma^\nu,\gamma^z]\right)  g\left(z_2 n_z\right) h\left(z_3 n_z\right)\right| B(P_B,\lambda=\frac{3}{2})\right\rangle 
=
- \lambda_{\varphi}\tilde \varphi^B \left(z_1,z_2,z_3,P_B^z\right) \Delta^\nu.
\end{aligned}
\end{equation}

Similar to Eq.~(\ref{eq:LCDA}), a quasi-DA in momentum space is defined as
\begin{equation}\label{eq:quasi-DA}
\begin{aligned}
& \widetilde \Phi^B\left(x_1, x_2,x_3,P^z_B,\mu \right)=
\int_{-\infty}^{+\infty} \frac{P_B^zd \,  z_1}{2 \pi}  \frac{P_B^zd \, z_2}{2 \pi}
\times  e^{-i x_1 P_B^z z_1-i x_2 P_B^z z_2} 
\times \widetilde \Phi^B_R\left(z_1, z_2,0,P^z_B,\mu\right),
\end{aligned}
\end{equation}
It should be noted that, in LaMET, the renormalization scheme for a quasi-DA and that of an LCDA may differ. Such differences will affect the UV behavior and should be compensated by the hard kernel  $\mathcal{C}$.

\subsection{One-loop perturbative results of quasi-DAs}

\begin{figure}[htb]
\includegraphics[width=0.6\textwidth]{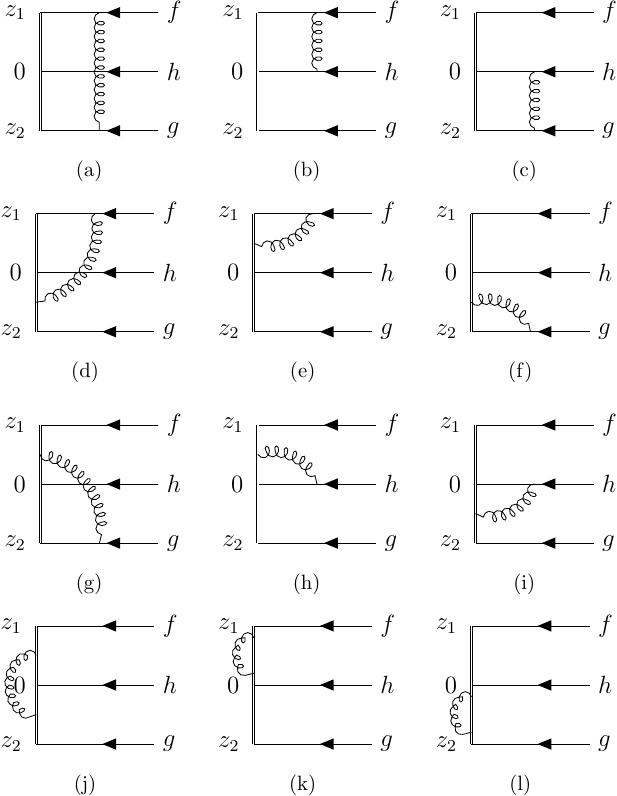}
\captionsetup{justification=raggedright,singlelinecheck=false}
\caption{One-loop corrections to quasi-DAs. The diagrams are classified based on the different types of divergences they have, corresponding to the manner of gluon attachment to quarks or Wilson lines. In the first line, which collects the quark-gluon-quark pattern, only IR divergences are present. The fourth line, displaying Wilson-line self-energy type diagrams, exhibits solely UV divergences. For the quark-gluon-Wilson-line exchange diagrams in the second and third lines, both UV and IR divergences manifest in the perturbative result.}
\label{Pspic}
\end{figure}

In this subsection, we present the perturbative results of quasi-DAs and the corresponding zero-momentum matrix elements up to the next-to-leading order in $\alpha_s$. Since the hard kernel represents the UV difference between the operator in the LCDA and that in the quasi-DA, the hadron state is irrelevant. Therefore, replacing the hadron state with a partonic state with the same quantum numbers remains valid for the sake of convenience in perturbative calculations. Thereby, we conduct a perturbative calculation of quasi-DAs by sandwiching the operators between the vacuum state $\langle0|$ and the lowest-order Fock state $| f_a(x_1 P) g_b(x_2 P) h_c(x_3 P)\rangle$.
For example, the matrix element $\widetilde{M}_V$ of a proton is defined as
\begin{eqnarray}
&&\widetilde{\mathcal{M}}_{V}(z_1,z_2,z_3,P^z,\mu) 
= \frac{\epsilon^{ijk}\epsilon^{abc}}{6} \left\langle 0\left|u_i^T\left(z_1\right)C \slashed n_z u_j\left(z_2\right)  d_k\left(z_3\right)\right| u_a(x_1 P) u_b(x_2 P) d_c(x_3 P)\right\rangle.
\end{eqnarray} 
Other structures can be obtained using the same method, with an appropriate choice of Fock states and Dirac matrices based on Eqs.~(\ref{quasi-def-1}, \ref{quasi-def-2}). The one-loop diagrams of quasi-DA are depicted in Fig.~(\ref{Pspic}).

Up to one-loop accuracy, the $\overline {\rm MS}$  renormalized spatial correlators are given as
\begin{align} \label{eq:Mperv} 
 &\widetilde {\mathcal{M}}_{V(A)}(z_1,z_2,0,P^z,\mu)= \notag
  \\& 
  \left\{1 
 + \frac{\alpha _s C_F}{\pi }\left(
 \frac{1}{2}  L_1^\text{UV}
 +
 \frac{1}{2}  L_2^\text{UV}
 +
 \frac{1}{2}  L_{12}^\text{UV}
 +\frac{3}{2}
 \right)\right\} 
 \widetilde {\mathcal{M}}_0 \left(z _1 , z _2 , 0 ,P^z,\mu\right) 
 \notag\\& 
-\frac{\alpha _s C_F}{8 \pi } \int_0^1 d \eta _1 \int_0^{1-\eta _1} d \eta _2  
 \notag\\& 
\times \left\{
\left (L_1^\text{IR}-1+\frac{1}{\epsilon_{\mathrm{IR}}}\right )
\widetilde {\mathcal{M}}_0 \left(\left(1-\eta _1\right) z _1 , z _2 , \eta _2 z _1 , P^z,\mu \right) 
+\left (L_2^\text{IR}-1+\frac{1}{\epsilon_{\mathrm{IR}}}\right )
\widetilde {\mathcal{M}}_0 \left(z _1 , \left(1-\eta _1\right) z _2 , \eta _2 z _2 , P^z,\mu \right)
\right.
\notag\\&
\left.+2 
\left (L_{12}^\text{IR}-3+\frac{1}{\epsilon_{\mathrm{IR}}}\right )
\widetilde {\mathcal{M}}_0 \left(\left(1-\eta _1\right) z _1+\eta _1 z _2 , \left(1-\eta _2\right) z _2+\eta _2 z _1 , 0 , P^z,\mu \right) \right\}
 \notag\\& 
-\frac{\alpha _s C_F}{4 \pi } \int_0^1 d \eta  
\times\left\{\widetilde {\mathcal{M}}_0 \left((1-\eta ) z _1+\eta z _2 , z_2 , 0 , P^z,\mu \right) 
\left\{
\left (L_{12}^\text{IR}+1+\frac{1}{\epsilon_{\mathrm{IR}}}\right )
\left(\frac{1-\eta }{\eta }\right)_+ 
+2 \left(\frac{\ln  \eta }{\eta }\right)_+ \right\}  \right.
\notag \\& 
+\widetilde {\mathcal{M}}_0 \left(z _1 , (1-\eta ) z _2+\eta z _1 , 0 , P^z,\mu \right) 
\left\{ 
\left (L_{12}^\text{IR}+1+\frac{1}{\epsilon_{\mathrm{IR}}}\right )
\left(\frac{1-\eta }{\eta }\right)_+
+2 \left(\frac{\ln  \eta}{\eta }\right)_+ 
\right\} 
 \notag\\& 
+\widetilde {\mathcal{M}}_0 \left((1-\eta ) z _1 , z _2 , 0 , P^z,\mu \right) 
\left\{
\left(
L_1^\text{IR}+1+\frac{1}{\epsilon_{\mathrm{IR}}}
\right)
\left(\frac{1-\eta }{\eta }\right)_+ 
+2 \left(\frac{\ln  \eta  }{\eta }\right)_+ \right\}   
\notag \\& 
+\widetilde {\mathcal{M}}_0 \left(z _1 , (1-\eta ) z _2 , 0 , P^z,\mu \right) 
\left\{
\left(L_2^\text{IR}+1+\frac{1}{\epsilon_{\mathrm{IR}}}
\right)
\left(\frac{1-\eta }{\eta }\right)_+ 
+2 \left(\frac{\ln  \eta  }{\eta }\right)_+ \right\}  
\notag\\& 
-\widetilde {\mathcal{M}}_0 \left(z _1 , z _2 , \eta  z _1 , P^z,\mu \right) 
\left\{ 
\left (L_1^\text{IR}+1+\frac{1}{\epsilon_{\mathrm{IR}}}\right )
\left(\frac{1-\eta }{\eta }\right)_+
+2 \left(\frac{\ln  \eta }{\eta }\right)_+\right\} 
 \notag\\ &
\left.-\widetilde {\mathcal{M}}_0 \left(z _1 , z _2 , \eta  z _2 , P^z,\mu \right) 
\left\{ 
\left (L_2^\text{IR}+1+\frac{1}{\epsilon_{\mathrm{IR}}}\right )
\left(\frac{1-\eta }{\eta }\right)_+
+2 \left(\frac{\ln  \eta }{\eta }\right)_+\right\}\right\},
 \end{align}     
 \begin{align} \label{eq:Mpert}
 &\widetilde {\mathcal{M}}_T(z_1,z_2,0,P^z,\mu)= \notag\\
  &\left\{1 
 + \frac{\alpha _s C_F}{\pi }\left(
 \frac{1}{2}  L_1^\text{UV}
 +
 \frac{1}{2}  L_2^\text{UV}
 +
 \frac{1}{2}  L_{12}^\text{UV}
 +\frac{3}{2}
 \right)\right\}  
 \widetilde {\mathcal{M}}_0 \left(z _1 , z _2 , 0 ,P^z,\mu\right) 
 \notag\\& 
-\frac{\alpha _s C_F}{8 \pi } \int_0^1 d \eta _1 \int_0^{1-\eta _1} d \eta _2  
 \notag\\& 
\times \left\{
\left (L_1^\text{IR}-1+\frac{1}{\epsilon_{\mathrm{IR}}}\right )
\widetilde {\mathcal{M}}_0 \left(\left(1-\eta _1\right) z _1 , z _2 , \eta _2 z _1 , P^z,\mu \right) 
+\left (L_2^\text{IR}-1+\frac{1}{\epsilon_{\mathrm{IR}}}\right )
\widetilde {\mathcal{M}}_0 \left(z _1 , \left(1-\eta _1\right) z _2 , \eta _2 z _2 , P^z,\mu \right)
 \right\}
 \notag\\& 
-\frac{\alpha _s C_F}{4 \pi } \int_0^1 d \eta  
\times\left\{\widetilde {\mathcal{M}}_0 \left((1-\eta ) z _1+\eta z _2 , z_2 , 0 , P^z,\mu \right) 
\left\{
\left (L_{12}^\text{IR}+1+\frac{1}{\epsilon_{\mathrm{IR}}}\right )
\left(\frac{1-\eta }{\eta }\right)_+ 
+2 \left(\frac{\ln  \eta }{\eta }\right)_+ \right\}  \right.
\notag \\& 
+\widetilde {\mathcal{M}}_0 \left(z _1 , (1-\eta ) z _2+\eta z _1 , 0 , P^z,\mu \right) 
\left\{ 
\left (L_{12}^\text{IR}+1+\frac{1}{\epsilon_{\mathrm{IR}}}\right )
\left(\frac{1-\eta }{\eta }\right)_+
+2 \left(\frac{\ln  \eta}{\eta }\right)_+ 
\right\} 
 \notag\\& 
+\widetilde {\mathcal{M}}_0 \left((1-\eta ) z _1 , z _2 , 0 , P^z,\mu \right) 
\left\{
\left(
L_1^\text{IR}+1+\frac{1}{\epsilon_{\mathrm{IR}}}
\right)
\left(\frac{1-\eta }{\eta }\right)_+ 
+2 \left(\frac{\ln  \eta  }{\eta }\right)_+ \right\}   
\notag \\& 
+\widetilde {\mathcal{M}}_0 \left(z _1 , (1-\eta ) z _2 , 0 , P^z,\mu \right) 
\left\{
\left(L_2^\text{IR}+1+\frac{1}{\epsilon_{\mathrm{IR}}}
\right)
\left(\frac{1-\eta }{\eta }\right)_+ 
+2 \left(\frac{\ln  \eta  }{\eta }\right)_+ \right\}  
\notag\\& 
-\widetilde {\mathcal{M}}_0 \left(z _1 , z _2 , \eta  z _1 , P^z,\mu \right) 
\left\{ 
\left (L_1^\text{IR}+1+\frac{1}{\epsilon_{\mathrm{IR}}}\right )
\left(\frac{1-\eta }{\eta }\right)_+
+2 \left(\frac{\ln  \eta }{\eta }\right)_+\right\} 
 \notag\\ &
\left.-\widetilde {\mathcal{M}}_0 \left(z _1 , z _2 , \eta  z _2 , P^z,\mu \right) 
\left\{ 
\left (L_2^\text{IR}+1+\frac{1}{\epsilon_{\mathrm{IR}}}\right )
\left(\frac{1-\eta }{\eta }\right)_+
+2 \left(\frac{\ln  \eta }{\eta }\right)_+\right\}\right\},
 \end{align}     
 \begin{align} \label{eq:Mperphi}
 &\widetilde {\mathcal{M}}_{\varphi}(z_1,z_2,0,P^z,\mu)= \notag
 \\& 
  \left\{1 
 + \frac{\alpha _s C_F}{\pi }\left(
 \frac{1}{2}  L_1^\text{UV}
 +
 \frac{1}{2}  L_2^\text{UV}
 +
 \frac{1}{2}  L_{12}^\text{UV}
 +\frac{3}{2}
 \right)\right\} 
 \widetilde {\mathcal{M}}_0 \left(z _1 , z _2 , 0 ,P^z,\mu\right) 
 \notag\\& 
-\frac{\alpha _s C_F}{4 \pi } \int_0^1 d \eta  
\times\left\{\widetilde {\mathcal{M}}_0 \left((1-\eta ) z _1+\eta z _2 , z_2 , 0 , P^z,\mu \right) 
\left\{
\left (L_{12}^\text{IR}+1+\frac{1}{\epsilon_{\mathrm{IR}}}\right )
\left(\frac{1-\eta }{\eta }\right)_+ 
+2 \left(\frac{\ln  \eta }{\eta }\right)_+ \right\}  \right.
\notag \\& 
+\widetilde {\mathcal{M}}_0 \left(z _1 , (1-\eta ) z _2+\eta z _1 , 0 , P^z,\mu \right) 
\left\{ 
\left (L_{12}^\text{IR}+1+\frac{1}{\epsilon_{\mathrm{IR}}}\right )
\left(\frac{1-\eta }{\eta }\right)_+
+2 \left(\frac{\ln  \eta}{\eta }\right)_+ 
\right\} 
 \notag\\& 
+\widetilde {\mathcal{M}}_0 \left((1-\eta ) z _1 , z _2 , 0 , P^z,\mu \right) 
\left\{
\left(
L_1^\text{IR}+1+\frac{1}{\epsilon_{\mathrm{IR}}}
\right)
\left(\frac{1-\eta }{\eta }\right)_+ 
+2 \left(\frac{\ln  \eta  }{\eta }\right)_+ \right\}   
\notag \\& 
+\widetilde {\mathcal{M}}_0 \left(z _1 , (1-\eta ) z _2 , 0 , P^z,\mu \right) 
\left\{
\left(L_2^\text{IR}+1+\frac{1}{\epsilon_{\mathrm{IR}}}
\right)
\left(\frac{1-\eta }{\eta }\right)_+ 
+2 \left(\frac{\ln  \eta  }{\eta }\right)_+ \right\}  
\notag\\& 
-\widetilde {\mathcal{M}}_0 \left(z _1 , z _2 , \eta  z _1 , P^z,\mu \right) 
\left\{ 
\left (L_1^\text{IR}+1+\frac{1}{\epsilon_{\mathrm{IR}}}\right )
\left(\frac{1-\eta }{\eta }\right)_+
+2 \left(\frac{\ln  \eta }{\eta }\right)_+\right\} 
 \notag\\ &
\left.-\widetilde {\mathcal{M}}_0 \left(z _1 , z _2 , \eta  z _2 , P^z,\mu \right) 
\left\{ 
\left (L_2^\text{IR}+1+\frac{1}{\epsilon_{\mathrm{IR}}}\right )
\left(\frac{1-\eta }{\eta }\right)_+
+2 \left(\frac{\ln  \eta }{\eta }\right)_+\right\}\right\}.
 \end{align}     
Here $\widetilde {\mathcal{M}}_0$ stands for tree-level matrix element. For example, the corresponding $\widetilde {\mathcal{M}}_0$ for $V$ is
\begin{equation}
\label{eq:2local}
\widetilde {\mathcal{M}}_{0}(z_1,z_2,0,P^z,\mu) = \sqrt{2} P^z e^{i x_1 P^z z_1+i x_2 P^z z_2} u_h(x_3 P),
\end{equation}
where $u_{h}(P)$ denotes the spinor of $h$ quark with momentum $P$.
The plus function is defined as 
\begin{eqnarray}
    \displaystyle\int_0^1 d u \left[ G(u) \right]_{+} F(u)=\displaystyle\int_0^1 d u G(u) [F(u)-F(0)].
\end{eqnarray}
Some abbreviations are used in Eqs.~(\ref{eq:Mperv}, \ref{eq:Mpert}, \ref{eq:Mperphi}):
\begin{eqnarray}
L_1^{\text{IR, UV}}=\ln \left(\displaystyle\frac{1}{4}\mu_{\text{IR, UV}} ^2 z_1^2 e^{2 \gamma_E }\right), 
L_2^{\text{IR,UV}}=\ln \left(\displaystyle\frac{1}{4}\mu_{\text{IR,UV}} ^2 z_2^2 e^{2 \gamma_E }\right), 
L_{12}^{\text{IR,UV}}=\ln\left(\displaystyle\frac{1}{4}\mu^2_{\text{IR,UV}}(z_1-z_2)^2 e^{2\gamma_E}\right). 
\end{eqnarray}
The one-loop results for $\tilde{T}$ and $\tilde{\varphi}$ are different from $\tilde{V}$ and $\tilde{A}$, due to the spin structure inherited in $\tilde{T}$ and $\tilde{\varphi}$. Explicitly, Fig.~1(a) vanishes for $\tilde{T}$, and Fig.~1(a-c) vanish for $\tilde{\varphi}$. We have verified that these results are consistent with the calculations performed in momentum space. 

Especially, there are two important cases in the one-loop results of spatial correlators. The first one is the local matrix element $\widetilde{M}(0,0,0,P^z,\mu)$, through which the spatial correlators can be normalized as
\begin{equation}\label{eq:mo}
\begin{aligned}
\widehat{\mathcal{M}}_{V,A,T,\varphi}\left(z _1 , z _2 , z_3=0 , P^z,\mu \right) 
= \frac{\widetilde{\mathcal{M}}_{V,A,T,\varphi}(z_1,z_2,0,P^z,\mu)}{\widetilde{\mathcal{M}}_{V,A,T,\varphi}(0,0,0,P^z,\mu)}.
\end{aligned}
\end{equation}
This normalization cancels the IR poles in the form of
$\displaystyle \frac{1}{\epsilon_{\rm IR}}$.

The one-loop zero-momentum matrix element, 
which will be used in the hybrid renormalization scheme, is presented as
    \begin{align}\label{zmm}
    \widehat {\mathcal{M}}_V(0,0,0,0,\mu)=\widehat {\mathcal{M}}_A(0,0,0,0,\mu)&=1+\frac{\alpha_s C_F}{2 \pi}(\frac{7}{8}L_1+\frac{7}{8}L_2+\frac{3}{4}L_{12}+4),
    \notag\\
    \widehat {\mathcal{M}}_T(0,0,0,0,\mu)&=1+\frac{\alpha_s C_F}{2 \pi}(\frac{7}{8}L_1+\frac{7}{8}L_2+\frac{1}{2}L_{12}+\frac{13}{4}),
    \notag\\
    \widehat {\mathcal{M}}_{\varphi}(0,0,0,0,\mu)&=1+\frac{\alpha_s C_F}{2 \pi}(\frac{7}{8}L_1+\frac{7}{8}L_2+\frac{1}{2}L_{12}+3).
    \end{align}
With this result, the UV logarithm behaviour of these perturbative matrix elements, as $z_i\to 0$, are explicit.

\section{hybrid renormalization and matching}
\label{sec:hybrid}

In LaMET, partons with infinite momentum can be approximately accessed  by a hadron possessing large yet finite momentum. Nonetheless, employing this approximation is not straightforward, primarily due to the existence of UV divergences. Addressing these divergences necessitates the utilization of conventional effective field theory methodologies such as matching and the renormalization group equation~\cite{Ji:2020ect}. The aim of this section is to introduce a suitable renormalization scheme capable of effectively renormalizing all spatial correlators of baryons. Subsequently, we will derive the corresponding hard kernels through the matching formula in Eq.~(\ref{matching}).

To begin, we present the challenges encountered in renormalizing spatial correlators within the framework of LaMET. To deal with the UV divergences inherent in spatial correlators, several renormalization schemes have been taken into consideration~\cite{Deng:2023csv,Han:2023xbl,Han:2023hgy}. A comprehensive overview of them will be provided in
this section. It becomes apparent that an effective and thorough renormalization scheme should cope with UV singular logarithms $\ln (\mu^2 z_i^2)$ alongside UV divergences. Subsequently, we present the hybrid renormalization scheme that serves the purpose and finally we obtain the hard kernels for $V,A,T$, and $\varphi$ through the matching.

\subsection{Difficulties in renormalizing spatial correlators}    
In the context of LaMET, spatial correlators play two different roles. Those with Fock states contribute to the perturbation computation of hard kernels, and those with hadron states serve as the non-perturbative inputs, simulated by Lattice QCD, to calculate LCDAs. It is evident that the validity of the factorization  of quasi-DAs in Eq.~(\ref{matching}) relies on the consistency of the renormalization scheme employed in both our perturbative calculations and Lattice QCD simulations. The first difficulty arises from the discrepancy between dimensional regularization, which facilitates easier perturbative calculations, and lattice regularization, which is inherent in Lattice QCD. 

The disparity between lattice simulation and perturbative calculation manifests in two key aspects. Firstly, Poincaré symmetry is broken by finite lattice spacing $a$ in Lattice QCD, resulting in power divergences and discrete effects within lattice matrix elements. Eliminating these effects becomes imperative for extrapolating lattice-simulated quasi-DAs to the continuum limit $a \to 0$.

The second distinction arises from the UV singular logarithms observed in the perturbative calculations in section III.C. They introduce additional logarithmic divergences when $z_1\to 0$, $z_2\to 0$ or $z_2-z_1\to 0$. On the other hand, lattice simulation yields a finite result. This discrepancy originates from the non-interchangeable limits: $a\to 0$ and $z \to 0$, {which are erroneously assumed in the composite local lattice operators}. It was proposed that a properly defined plus function could rectify this issue~\cite{Izubuchi:2018srq}.
However, these logarithmic terms correspond to slowly decaying terms \(\sim \frac{1}{|x'_1|}\) and \(\frac{1}{|x'_2|}\) in the hard kernel, thereby increasing the difficulty of numerical calculations in preserving the normalization, as the logarithms converge slowly.
An alternative and more efficient strategy is to directly eliminate the UV singular logarithms \(\ln (\mu^2 z_i^2)\) at short distances.

Dealing with these issues becomes  more challenging when considering baryons. The coordinates of baryon LCDAs form a two-dimensional distribution while baryons LCDAs contains three types of UV singular logarithms. 
Analyzing a baryon spatial correlator necessitates the examination of regions that simultaneously encompass both short and large distance scales. Examples include scenarios such as \(z_1 \rightarrow 0\) and \(z_2 \sim 1/\Lambda_{\rm QCD}\), \(z_2 \rightarrow 0\) and \(z_1 \sim 1/\Lambda_{\rm QCD}\), or \(z_1-z_2 \rightarrow 0\), \(z_1 \sim 1/\Lambda_{\rm QCD}\), and \(z_2 \sim 1/\Lambda_{\rm QCD}\) (excluding the one around \(z_1 \sim -z_2\)), as depicted by $HSI$-$HSIII$ regions in Fig.~\ref{pic:Renorm}. In these regions, it becomes imperative to eliminate UV singular logarithms while preserving the underlying physics associated with the other non-perturbative 
 distance scale. 
If the logarithmic divergences associated with different scales, \(z_1\), \(z_2\), and \(z_1-z_2\), can be factorized independently, the UV divergences pertaining to these scales can be multiplicatively renormalized separately. 
Fortunately, the one-loop results in Eq.~(\ref{zmm}) demonstrates the factorization of UV singular logarithms. 
However, a comprehensive all-order proof remains a subject for future investigations.

\subsection{Renormalization schemes}
To devise a proper renormalization scheme that addresses the issues outlined in the preceding subsection, a thorough examination of suitable renormalization schemes is warranted. The following list provides an overview of potential options:

\begin{itemize}
\item \textbf{$\overline{\mathrm{MS}}$ scheme:}  The $\overline{\mathrm{MS}}$ scheme is one of the most prevalent renormalization scheme. 
However, performing direct matching in the $\overline{\mathrm{MS}}$ scheme faces challenges in practical calculations due to inconsistencies between lattice matrix elements and perturbative matrix elements at short distances.
While lattice matrix elements remain finite as \(z_1 \rightarrow 0\), \(z_2 \rightarrow 0\), or \(z_1-z_2 \rightarrow 0\), perturbative matrix elements exhibit UV logarithmic divergences in the forms of  \(\ln(\mu^2 z_1^2)\), \(\ln(\mu^2 z_2^2)\), and \(\ln(\mu^2 (z_1-z_2)^2)\).

\item \textbf{RI/MOM:} A widely employed non-perturbative renormalization method for Lattice QCD matrix elements, designed to eliminate the logarithms, is the standard regularization invariant momentum subtraction  (RI/MOM)~\cite{Martinelli:1994ty}. 
In the RI/MOM approach, matrix elements are computed with an off-shell partonic state with momentum satisfying \(-P^2\gg \Lambda_{\text{QCD}}^2\).
The corresponding counterterm up to the next-leading order in $\alpha_s$ for the distribution amplitude $A$ of light baryons in this scheme has been derived~\cite{Deng:2023csv}.
Despite the theoretical merits of the RI/MOM scheme, 
detailed scrutiny reveals that this method may introduce potential non-perturbative effects to non-local lattice matrix elements, particularly through IR logarithms such as $\ln( \mu^2z_i^2)$.

\item \textbf{Ratio scheme:} The resolution of logarithmic divergences issue can be solved through the application of the ratio scheme~\cite{Orginos:2017kos,Radyushkin:2017cyf,Radyushkin:2017lvu}. 
By dividing the perturbative matrix element by an appropriate zero-momentum matrix element, this scheme addressed several challenges.
Due to the multiplicative renormalizability of the spatial correlators, the UV $\ln(\mu^2 z_i^2)$ terms in the numerator and denominator are equal by construction and cancel out. 
Besides, it enables the cancellation of part of the discretization effects present in lattice matrix elements, allowing for the continuum limit to be taken.
However, the effectiveness relies on the validity of the Euclidean operator product expansion (OPE) and is only applicable to correlators at short distances. 
If the zero-momentum matrix element contains physics at scale $\Lambda_{\rm QCD}$, the IR structure of the renormalized matrix element may be altered. Consequently, the choice of zero-momentum matrix elements in the short-distance region as denominators is a requirement.

\item \textbf{Self-renormalization scheme:} The self-renormalization scheme, as advocated in Ref.~\cite{LatticePartonCollaborationLPC:2021xdx}, has emerged in recent years as a novel renormalization approach. 
In this scheme, a renormalization factor \(Z_R\), encompassing all typical divergences, especially the linear divergences, and discretization errors of Lattice QCD, is defined. 
Utilizing this renormalization factor enables the conversion of lattice matrix elements to continuous matrix elements with the non-perturbative physics intact. The UV singular logarithms  still remain after the application of self-renormalization.

\end{itemize}

These listed renormalization schemes cannot solely eliminate the UV divergence as well as the UV singular logarithms when $z_i$ approaches 0. A properly defined renormalization scheme should be built up as a combination of this renormalization schemes.

\subsection{Hybrid renormalization scheme}
        
To address the challenges outlined earlier, the hybrid renormalization scheme developed in Ref.~\cite{Han:2023xbl} will be employed.    
Based on the discussion mentioned before, the hybrid renormalization  scheme adheres to several key principles, as outlined in Ref.~\cite{LatticePartonCollaborationLPC:2021xdx}:
\begin{itemize}
    \item  \textbf{Elimination of Singular Logarithmic Terms}: The hybrid renormalization scheme aims to eliminate all UV singular logarithms in perturbative matrix elements, including \(\ln(\mu^2 z_1^2)\), \(\ln(\mu^2 z_2^2)\), and \(\ln(\mu^2(z_1-z_2)^2)\).
    \item \textbf{Avoidance of Uncontrollable Effects at Large Distances}: The scheme should avoid introducing uncontrollable non-perturbative effects.
    \item  \textbf{Continuity and Simplicity}: The hybrid renormalization  scheme is designed to keep the renormalized matrix element continuous and the method as simple as possible. This ensures both the coherence of the renormalization process and the practicality of its implementation.

\end{itemize}

The outline of the hybrid renormalization  scheme is constructed upon the aforementioned principles:
To remove all the typical divergences in Lattice QCD, the self-renormalization method is suggested. 
Through self-renormalization, these divergences can be eliminated without alternating the physics at scale $\Lambda_{\rm{QCD}}$.
To eliminate the logarithms $\ln(\mu^2z_i^2)$, which undermines the validation of OPE, RI/MOM scheme or ratio scheme is recommended.
In this paper, the ratio scheme is employed. 
Explicitly, the zero-momentum matrix element, whose coordinates may be different from the matrix element that needs to be renormalized, is chosen as the denominator for the ratio scheme.
This choice of its coordinates plays a pivotal role because the applicability of the ratio scheme is limited to $z_i\ll 1/\Lambda_{\rm{QCD}}$. Otherwise the non-perturbative structure of the bare matrix element is altered by the ratio scheme.
Thus a proper choice is needed which not only eliminates the UV singular logarithms but also leaves physics at scale $\Lambda_{QCD}$ intact.
Explicitly, the distribution needs to be partitioned into different parts, each requiring a choice of appropriate zero-momentum matrix elements.

\subsubsection{Employing the self-renormalization}

Based on this blueprint, we employ the self-renormlization to the lattice spatial correlator \(\hat{M}(z_1, z_2, 0, P^z, a)\), where the normalization scale $\mu$ is replaced by the lattice spacing $a$, through a renormalization factor \(Z_R(z_1, z_2, a, \mu)\).
Here, the small hat in \(\hat{M}(z_1, z_2, 0, P^z, a)\) stands for lattice results, which are normalized similarly to the perturbative case, as shown in Eq.~(\ref{eq:mo}).
Ref.~\cite{LatticePartonCollaborationLPC:2021xdx} indicates that the renormalization factor is subject to an asymptotic expansion with respect to \(a\), encompassing both power and logarithmic dependencies
\begin{equation}\label{eq:ZRm0}
Z_{R}(z_1,z_2,a,\mu) = \exp\Big[\left(\frac{k}{a \ln[a \Lambda_{\rm QCD}]} 
- m_{0}\right) \tilde{z}+\frac{\gamma_0}{b_0} \ln \bigg[\frac{\ln [1 /(a \Lambda_{\rm QCD})]}{\ln [\mu / \Lambda_{\rm \overline{MS}}]}\bigg]+\ln \left[1+\frac{d}{\ln (a \Lambda_{\rm QCD})}\right] + f(z_1,z_2)a \Big]\ ,
\end{equation}    
where $\displaystyle\left(\frac{k}{a \ln[a \Lambda_{\rm QCD}]} 
- m_{0}\right) \tilde{z}$ represents the linear divergences in lattice spatial correlators~\cite{Chen:2016fxx,Ji:2017oey,Ishikawa:2017faj,Green:2017xeu, Ji:2020brr} and $m_0$ is the mass renormalization parameter~\cite{Ji:1995tm,Beneke:1998ui,Bauer:2011ws,Bali:2013pla,Zhang:2023bxs}, $\displaystyle\frac{\gamma_0}{b_0} \ln \bigg[\frac{\ln [1 /(a \Lambda_{\rm QCD})]}{\ln [\mu / \Lambda_{\rm \overline{MS}}]}\bigg]+\ln \left[1+\frac{d}{\ln (a \Lambda_{\rm QCD})}\right]$ contains the logarithmic divergences.  $f(z_1,z_2)a$ (or $f(z_1,z_2)a^2$) stands for the discretization effect. $\tilde{z}$ is the effective length for the linear divergence, which is defined as follows
\begin{eqnarray}
\tilde{z} = \left\{
        \begin{array}{ll}
            |z_1-z_2|, & \quad z_1 z_2 < 0 \\
            {\rm max}\left(|z_1|,|z_2|\right), & \quad z_1 z_2 \geq 0.
        \end{array}
    \right.
\end{eqnarray}
The leading-order (LO) QCD $\beta$ function $\displaystyle b_0=\frac{11 C_A - 2 n_f}{6\pi}$ satisfies $\displaystyle\frac{d \alpha_s}{d \ln\mu} = - b_0 \alpha_s^2$. The perturbative zero-momentum matrix element in $\overline{\rm MS}$ scheme, such as $\widehat{\mathcal{M}}_A$, satisfies the following renormalization group equation
\begin{equation}\label{eq:RGM0}
\frac{d \ln[\widehat{\mathcal{M}}_A\left(z_1, z_2, 0, 0, \mu\right)]}{d \ln{\mu}} = \gamma,
\end{equation}
where $\gamma=\sum_{i=0}^\infty \gamma_i \alpha_s^{i+1}$. 
The leading anomalous dimension $\displaystyle\gamma_0=\frac{C_F}{2\pi}\left(5-\frac{7}{4}\delta_{z_1,0}-\frac{7}{4}\delta_{z_2,0}-\frac{3}{2}\delta_{z_1-z_2,0}\right)$ is scheme independent and, thus, can be applied to the renormalization factor Eq.~(\ref{eq:ZRm0}) of lattice matrix elements. It involves the quark-link interaction as well as the evolution effect of the local operator. There are subtraction terms since the UV fluctuation is frozen on lattice for a distance to be zero. $\Lambda_{\rm \overline{MS}}$ is the RG invariant scale for the LO running coupling, which is 0.142 GeV for $n_f=3$, 0.119 GeV for $n_f=4$ and 0.087 GeV for $n_f=5$, determined based on the method in~\cite{Karbstein:2018mzo}.  

The parameters $k$, $\Lambda_{\rm QCD}$, $f(z_1,z_2)$, $m_0$ and $d$ are extracted through the fit. The fit procedure is~\cite{LatticePartonCollaborationLPC:2021xdx}:\\ 
1) fit the $a$ dependence in $\hat{M}\left(z_1, z_2, 0, 0, a\right)$ to extract the global parameters $k$ and $\Lambda_{\rm QCD}$ as well as the discretization effect $f(z_1,z_2)$:
\begin{align}
&\hat{M}\left(z_1, z_2, 0, 0, a\right) \notag
\\&
= \exp\Big[\frac{k}{a \ln[a \Lambda_{\rm QCD}]}\tilde{z}+g(z_1,z_2,d)+\frac{\gamma_0}{b_0} \ln \bigg[\frac{\ln [1 /(a \Lambda_{\rm QCD})]}{\ln [\mu / \Lambda_{\rm \overline{MS}}]}\bigg]+\ln \left[1+\frac{d}{\ln (a \Lambda_{\rm QCD})}\right] + f(z_1,z_2)a \Big]\ ,
\end{align}    
where $g(z_1,z_2,d)$ contains the non-perturbative intrinsic $z_1,z_2$ dependence, which is also extracted through the fit. 
It should be mentioned that the dependence of $g(z_1,z_2,d)$ on the global parameter $d$ stems from the choice of $d$ during the fitting.
\\
2) extract $m_0$ and $d$ through requiring the renormalized matrix element equal to the perturbative matrix element at short distances ($a < z_1,z_2 \ll 1/\Lambda_{\rm QCD}$)
\begin{equation}
\frac{\hat{M}\left(z_1, z_2, 0, 0, a\right)}{Z_{R}(z_1,z_2,a,\mu)}=\exp\Big[ g(z_1,z_2,d) + m_0 \tilde{z} \Big]=\widehat{\mathcal{M}}\left(z_1, z_2, 0, 0, \mu\right),
\end{equation}
where $\widehat{M}$ is defined in Eq.~(\ref{zmm}) for $V,A,T$, and $\varphi$, respectively.

Thus one can define the renormalized lattice matrix element in $\overline{\rm MS}$ scheme for the whole range as the ratio of the normalized lattice matrix element to the renormalization factor Eq.~(\ref{eq:ZRm0})
\begin{equation}\label{eq:latinMSbar}
\hat{M}_{\overline{\rm MS}}\left(z_1, z_2, 0, P^z, \mu\right) = \frac{\hat{M}\left(z_1, z_2, 0, P^z, a\right)}{Z_{R}(z_1,z_2,a,\mu)},
\end{equation}
where the renormalization factor $Z_{R}$, though extracted from the zero-momentum matrix element, can be applied to the large momentum matrix element since the renormalization is independent of the external states.

\subsubsection{Performing the ratio scheme}
After applying the self-renormalization, the next step involves implementing the ratio scheme by dividing the matrix element by a zero-momentum matrix element.
Subsequently, it is necessary to categorize the distribution into different regions and select appropriate zero-momentum matrix elements for each region. 
The separation criteria can be defined as follows:
\begin{itemize}

    \item \textbf{$H$ Region}: All the UV singular logarithms need to be eliminated.
        \item  \textbf{$HSI$-$HSIII$ Regions}: The UV singular logarithms $\ln(\mu^2 z_1^2)$, $\ln(\mu^2 z_2^2)$ and $\ln(\mu^2 (z_1-z_2)^2)$ need to be eliminated in HSI, HSII and HSIII, respectively.
            \item  \textbf{$HSIV$ Regions}: No UV singular logarithms need to be eliminated.
    \item \textbf{$S$ Regions}: No UV singular logarithms need to be eliminated.

\end{itemize}

\begin{figure}[http]
\centering
\includegraphics[width=1.0\textwidth]{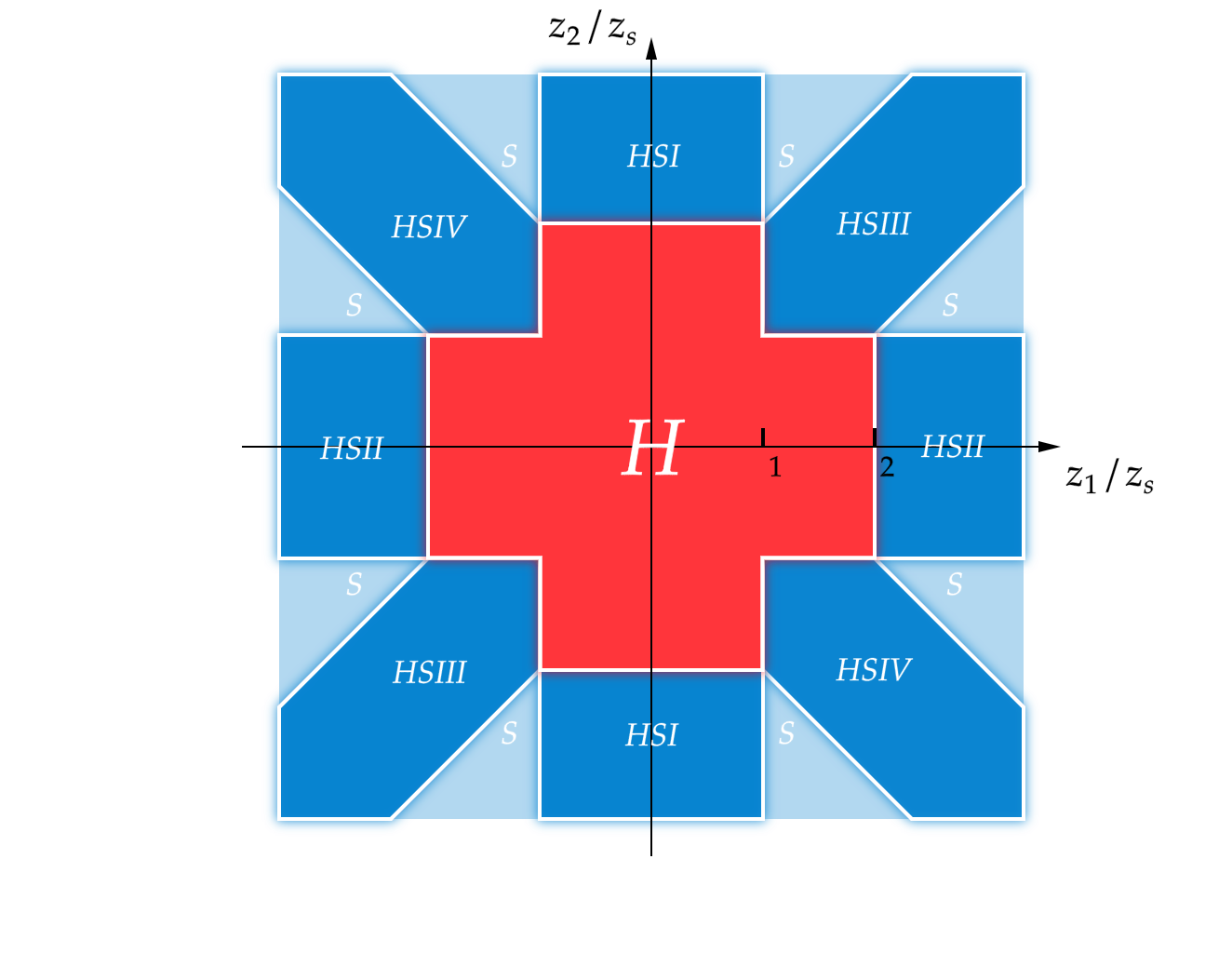}
\captionsetup{justification=raggedright,singlelinecheck=false}
\caption{
The regions separated by the scale $z_s$ for the application of the ratio scheme.
In the ratio scheme, the zero-momentum matrix elements are chosen as follows: 
the matrix elements in the $H$ region are divided by the zero-momentum matrix elements at the same location; 
the matrix elements in the $HSI-HSIV$ regions are divided by the zero-momentum matrix elements on their boundaries against the H region, which contains the same $z_1,z_2,z_1-z_2$ and $z_1+z_2$ accordingly; 
the matrix elements in the $S$ region are divided by the zero-momentum matrix element at the corresponding vertex of the $S$ regions.
}
\label{pic:Renorm}
\end{figure}

 This division of regions elaborates the typical large logarithms which should be eliminated.
            
\begin{itemize}
\item For $|z_1|<z_s$, $|z_2|<z_s$ and $|z_1-z_2|<z_s$,
which corresponds to the $H$ region in Fig.~\ref{pic:Renorm}, one introduces the ratio scheme on the normalized lattice matrix elements
\begin{equation}
 \begin{aligned}
\frac{\hat{M}_{\overline{\rm MS}}\left(z_1, z_2, 0, P^z,\mu\right)}{\hat{M}_{\overline{\rm MS}}\left(z_1, z_2, 0, 0,\mu\right)} \left(\theta(2z_s-|z_1|)\theta(z_s-|z_2|)+\theta(z_s-|z_1|)\theta(|z_2|-z_s)\theta(2z_s-|z_2|)\right),
\end{aligned}
\end{equation}      
where $z_s$ is the length scale under which $ln (\mu^2z_i^2)$ needs to be eliminated. The typical choice of $z_s$ satisfies $a \ll 2z_s \ll 1/\Lambda_{\rm QCD}$. So no extra non-perturbative effects are introduced. The ratio can be written with the renormalized lattice matrix element in $\overline{\rm MS}$ scheme Eq.~(\ref{eq:latinMSbar}) since the renormalization factor $Z_R$ is independent of momentum $P^z$. 

\item For $|z_1|<z_s$ and $|z_2|>2z_s$, which are shown as the $HSI$ regions in Fig.~\ref{pic:Renorm},
the zero-momentum matrix elements are selected by projecting the point along the y-axis to the boundaries of the $HSI$ and $H$ regions.

\begin{equation}
\begin{aligned}
\frac{\hat{M}_{\overline{\rm MS}}\left(z_1, z_2, 0, P^z, \mu\right)}{\hat{M}_{\overline{\rm MS}}\left(z_1, {\rm sign}(z_2)2z_s, 0, 0, \mu\right)}\theta(z_s-|z_1|)\theta(|z_2|-2z_s),
\end{aligned}
\end{equation}    
The zero-momentum matrix element in the denominator $\hat{M}\left(z_1, {\rm sign}(z_2)2z_s, 0, 0, a\right)$ is crucial in canceling the $\ln(z_1^2)$ in the perturbative matrix element when we deduce the hard kernel. 
No extra non-perturbative effect is introduced in the denominator where the $z_2$ dependence is truncated at ${\rm sign}(z_2)2z_s$.

\item For $|z_1|>2z_s$ and $|z_2|<z_s$, which are the $HSII$ regions in Fig.~\ref{pic:Renorm}, one follows the similar strategy, 
\begin{equation}
\frac{\hat{M}_{\overline{\rm MS}}\left(z_1, z_2, 0, P^z, \mu\right)}{\hat{M}_{\overline{\rm MS}}\left({\rm sign}(z_1)2z_s, z_2, 0, 0, \mu\right)}\theta(|z_1|-2z_s)\theta(z_s-|z_2|).
\end{equation}

\item For $|z_1|>z_s$, $|z_2|>z_s$ and $|z_1-z_2|<z_s$,  which are the $HSIII$ regions around $z_1 \sim z_2$ in Fig.~\ref{pic:Renorm}, one takes
\begin{equation}
\frac{\hat{M}_{\overline{\rm MS}}\left(z_1, z_2, 0, P^z, \mu\right)}{\hat{M}_{\overline{\rm MS}}\left(z_1^*, z_2^*, 0, 0, \mu\right)}\theta(|z_1|-z_s)\theta(|z_2|-z_s)\theta(z_s-|z_1-z_2|),
\end{equation}
where $z_1^*=z_s+(z_1-z_2)\theta(z_1-z_2)$ and $z_2^*=z_s+(z_2-z_1)\theta(z_2-z_1)$. The coordinate choices $z_1^*$ and  $z_2^*$ apply for $z_1<0$ and $z_2<0$ as well because the zero-momentum matrix element is invariant under the exchange $z_1 \leftrightarrow -z_2$. The zero-momentum matrix element in the denominator will be crucial in canceling the $\ln(\mu^2(z_1-z_2)^2)$ in the perturbative matrix element. 

\item For $|z_1|>z_s$, $|z_2|>z_s$ and $|z_1 + z_2|<z_s$, which are the $HSIV$ regions around $z_1 \sim -z_2$ in Fig.~\ref{pic:Renorm}, one can introduce the ratio as well
\begin{equation}
\frac{\hat{M}_{\overline{\rm MS}}\left(z_1, z_2, 0, P^z, \mu\right)}{\hat{M}_{\overline{\rm MS}}\left(z_1^{**}, z_2^{**}, 0, 0, \mu\right)}\theta(|z_1|-z_s)\theta(|z_2|-z_s)\theta(z_s-|z_1+z_2|),
\end{equation}
where $z_1^{**}=z_s+(z_1+z_2)\theta(z_1+z_2)$ and $z_2^{**}=-z_s+(z_2+z_1)\theta(-z_2-z_1)$. This step is for continuity and simplicity. 

\item Finally, for $|z_1|>z_s$, $|z_2|>z_s$, $|z_1-z_2|>z_s$ and $|z_1+z_2|>z_s$, namely the $S$ regions,
one can choose the zero-momentum matrix elements lie in the corresponding vertical and give 
\begin{equation}
\begin{aligned}
\frac{\hat{M}_{\overline{\rm MS}}\left(z_1, z_2, 0, P^z, \mu\right)}{\hat{M}_{\overline{\rm MS}}\left({\rm sign}(z_1)z_s, {\rm sign}(z_2)2z_s, 0, 0, \mu\right)}\theta(|z_1|-z_s)\theta(|z_2|-z_s)\theta(|z_1-z_2|-z_s)\theta(|z_1+z_2|-z_s).
\end{aligned}
\end{equation}    
\end{itemize}

Finally, this renormalization procedure can be extrapolated and then corresponding results can be given.
To conclude, the hybrid renormalized matrix element is
\begin{align}\label{eq:Mhybrid}
&\hat{M}_{H}(z_1,z_2,0,P^z) = \frac{\hat{M}_{\overline{\rm MS}}\left(z_1, z_2, 0, P^z,\mu\right)}{\hat{M}_{\overline{\rm MS}}\left(z_1, z_2, 0, 0,\mu\right)} \left(\theta(2z_s-|z_1|)\theta(z_s-|z_2|)+\theta(z_s-|z_1|)\theta(|z_2|-z_s)\theta(2z_s-|z_2|)\right) 
\notag\\
&+\frac{\hat{M}_{\overline{\rm MS}}\left(z_1, z_2, 0, P^z, \mu\right)}{\hat{M}_{\overline{\rm MS}}\left(z_1, {\rm sign}(z_2)2z_s, 0, 0, \mu\right)}\theta(z_s-|z_1|)\theta(|z_2|-2z_s)+\frac{\hat{M}_{\overline{\rm MS}}\left(z_1, z_2, 0, P^z, \mu\right)}{\hat{M}_{\overline{\rm MS}}\left({\rm sign}(z_1)2z_s, z_2, 0, 0, \mu\right)}\theta(|z_1|-2z_s)\theta(z_s-|z_2|) \notag\\
&+\frac{\hat{M}_{\overline{\rm MS}}\left(z_1, z_2, 0, P^z, \mu\right)}{\hat{M}_{\overline{\rm MS}}\left(z_s+(z_1-z_2)\theta(z_1-z_2), z_s+(z_2-z_1)\theta(z_2-z_1), 0, 0, \mu\right)}\theta(|z_1|-z_s)\theta(|z_2|-z_s)\theta(z_s-|z_1-z_2|) 
\notag\\
&+ \frac{\hat{M}_{\overline{\rm MS}}\left(z_1, z_2, 0, P^z, \mu\right)}{\hat{M}_{\overline{\rm MS}}\left(z_s+(z_1+z_2)\theta(z_1+z_2), -z_s+(z_2+z_1)\theta(-z_2-z_1), 0, 0, \mu\right)}\theta(|z_1|-z_s)\theta(|z_2|-z_s)\theta(z_s-|z_1+z_2|) 
\notag\\
&+\frac{\hat{M}_{\overline{\rm MS}}\left(z_1, z_2, 0, P^z, \mu\right)}{\hat{M}_{\overline{\rm MS}}\left({\rm sign}(z_1)z_s, {\rm sign}(z_2)2z_s, 0, 0, \mu\right)}\theta(|z_1|-z_s)\theta(|z_2|-z_s)\theta(|z_1-z_2|-z_s)\theta(|z_1+z_2|-z_s),
\end{align}    
where $\hat{M}_{\overline{\rm MS}}$ is the renormalized lattice matrix element defined in Eq.~(\ref{eq:latinMSbar}). 
Then the hybrid renormalization scheme quasi-DA can be obtained through the Fourier transformation
\begin{equation}
    \tilde \Phi_{H} (x_1,x_2,P^z,\mu) = \int_{-\infty}^{+\infty} \frac{ P^z d z_1}{2\pi} \frac{P^z d z_2}{2\pi} e^{-i x_1 P^z z_1- i x_2 P^z z_2} \hat{M}_{H}\left(z_1,z_2,0,P^z,\mu\right).
    \end{equation}

The hybrid renormalization  scheme can be seen as a refinement of the $\overline{\rm MS}$ scheme, particularly at short distances. 
The self-renormalization provides equivalent renormalized results in Lattice QCD compared to the $\overline{\rm MS}$ scheme in perturbative calculations.
By applying the ratio scheme in the short-distance regime, part of the discretization effects in the lattice matrix elements is removed, and singular logarithmic terms are canceled in the perturbative matrix elements.
As a result, the lattice matrix elements exhibit improved consistency with the continuum scheme within the hybrid renormalization  scheme. 
This facilitates a more straightforward preservation of normalization.

\subsection{Matching and hard kernel}

To obtain the LCDAs of the lowest-lying octet and decuplet baryons, we proceed by extracting them from the matching relations with the Fock state: 
\begin{equation}\label{Fock-mai}
\begin{aligned}
&\widetilde \Psi_{H}^{V,A,T,\varphi} (x_1,x_2,P^z,\mu) = 
\int d y_1 d y_2 \mathcal{C}_{V,A,T,\varphi}(x_1,x_2,y_1,y_2,P^z,\mu) \Psi_{\overline{\rm MS}}^{V,A,T,\varphi} (y_1,y_2,\mu),
\end{aligned}
\end{equation}
where $\Psi_{\overline{\rm MS}}^{V,A,T,\varphi}$ represent the perturbative LCDAs with Fock state renormalized under the $\overline{\rm{MS}}$ scheme, and $\widetilde \Psi_{H}^{V,A,T,\varphi}$ denote the perturbative quasi-DAs with Fock state renormalized under the hybrid renormalization scheme.
$\mathcal{C}_{{V,A,T,\varphi}}(x_1,x_2,y_1,y_2,P^z,\mu)$ stand for the hard kernels in the hybrid renormalization scheme.
Once the hard kernels are obtained, along with the quasi-DAs $\tilde{\Phi}_H^B$ provided by Lattice QCD, we can directly extract the LCDAs through matching with the hadron states:
\begin{equation}
\begin{aligned}
&\tilde \Phi_{H}^B (x_1,x_2,P^z,\mu) = 
\int d y_1 d y_2 \mathcal{C}_{{V,A,T,\varphi}}(x_1,x_2,y_1,y_2,P^z,\mu) \Phi_{\overline{\rm MS}}^B (y_1,y_2,\mu).
\end{aligned}
\end{equation}

In this subsection, we will start by presenting the one-loop perturbative results of lightcone correlators. Following that, we will introduce the one-loop perturbative results of spatial correlators within the hybrid renormalization scheme. Finally, we will provide all the necessary hard kernels required for obtaining the LCDAs of octet and decuplet baryons.

\subsubsection{One-loop perturbative results of lightcone correlators}
To carry out the matching procedure at the perturbative level, corresponding perturbative lightcone results are also required, necessitating the definition of a lightcone correlator. 
For instance, for the distribution amplitude $V$, it can be defined as:
\begin{eqnarray}
&&{\mathcal{I}}_{V}(z_1 n\cdot P,z_2 n\cdot P,z_3 n\cdot P,\mu) 
= \frac{\epsilon^{ijk}\epsilon^{abc}}{6} \left\langle 0\left|u_i^T\left(z_1\right)C \slashed n u_j\left(z_2\right)  d_k\left(z_3\right)\right| u_a(x_1 P) u_b(x_2 P) d_c(x_3 P)\right\rangle.
\end{eqnarray} 
\label{eq:sec-hybrid}
Other structures can be obtained using the same method, with an appropriate choice of Fock state and Dirac matrices based on Eqs.~(\ref{lc-oct-de}, \ref{lc-dec-de}). 
Unlike the spatial correlator case, the one-loop diagrams of the lightcone correlator do not include Wilson-line self-energy type diagrams.
Up to one-loop accuracy, the $\overline {\rm MS}$  renormalized lightcone correlators are given as
\begin{align} \label{light-va}
& \mathcal{I}_{V(A)}(\nu_1,\nu_2,0,\mu) = \mathcal{I}_0 \left(\nu_1 , \nu_2 , 0 ,\mu \right) 
 \notag\\& 
-\frac{\alpha _s C_F}{8 \pi } \frac{1}{\epsilon_{\text{IR}}}
\left\{  \int_0^1 d \eta _1 \int_0^{1-\eta _1} d \eta _2  \right.
\displaystyle\left[2\mathcal{I}_0 \left(\left(1-\eta _1\right) \nu_1+\eta _1 \nu_2 , \left(1-\eta _2\right) \nu_2+\eta _2 \nu_1 , 0 , \mu \right) \right.
\notag\\&
+ \mathcal{I}_0 \left(\left(1-\eta _1\right) \nu_1 , \nu_2 , \eta _2 \nu_1 , \mu \right)
\left.
+ \mathcal{I}_0 \left(\nu_1 , \left(1-\eta _1\right) \nu_2 , \eta _2 \nu_2 , \mu \right) \right]
\notag\\& 
+2 \int_0^1 d \eta  \left(\frac{1-\eta }{\eta }\right)_+ 
\left \{  \left(\mathcal{I}_0 \left((1-\eta ) \nu_1+\eta  \nu_2 , \nu_2 , 0 , \mu \right)+
\mathcal{I}_0 \left(\nu_1 , (1-\eta ) \nu_2+\eta  \nu_1 , 0 , \mu \right) \right) \right.
\notag\\& 
\left.\left.
 + 
 \left(\mathcal{I}_0 \left((1-\eta ) \nu_1 , \nu_2 , 0 , \mu \right)+\mathcal{I}_0 \left(\nu_1 , \nu_2 , \eta  \nu_1 , \mu \right)\right)
+
\left(
\mathcal{I}_0 \left(\nu_1 , (1-\eta ) \nu_2 , 0 , \mu 
\right)
+\mathcal{I}_0 \left(\nu_1 , \nu_2 , \eta  \nu_2 , \mu \right)
\right)
\right \} \right \} ,
\end{align} 
\begin{align} \label{light-t}
& \mathcal{I}_T(\nu_1,\nu_2,0,\mu) = \mathcal{I}_0 \left(\nu_1 , \nu_2 , 0 ,\mu \right) 
 \notag\\& 
-\frac{\alpha _s C_F}{8 \pi } \frac{1}{\epsilon_{\text{IR}}}
\left\{ 
\int_0^1 d \eta _1 \int_0^{1-\eta _1} d \eta _2  \right.
\displaystyle\left[\mathcal{I}_0 \left(\left(1-\eta _1\right) \nu_1 , \nu_2 , \eta _2 \nu_1 , \mu \right)
\left.
+ \mathcal{I}_0 \left(z_1 , \left(1-\eta _1\right) \nu_2 , \eta _2 \nu_2 , \mu \right) \right]\right.
\notag\\& 
+2 \int_0^1 d \eta  \left(\frac{1-\eta }{\eta }\right)_+ 
\left \{  \left(\mathcal{I}_0 \left((1-\eta ) \nu_1+\eta  \nu_2 , \nu_2 , 0 , \mu \right)+
\mathcal{I}_0 \left(\nu_1 , (1-\eta ) \nu_2+\eta  \nu_1 , 0 , \mu \right) \right) \right.
\notag\\& 
\left.\left.
 + 
 \left(\mathcal{I}_0 \left((1-\eta ) \nu_1 , \nu_2 , 0 , \mu \right)+\mathcal{I}_0 \left(\nu_1 , \nu_2 , \eta  \nu_1 , \mu \right)\right)
+
\left(
\mathcal{I}_0 \left(\nu_1 , (1-\eta ) \nu_2 , 0 , \mu 
\right)
+\mathcal{I}_0 \left(\nu_1 , \nu_2 , \eta  \nu_2 , \mu \right)
\right)
\right \} \right \} ,
\end{align} 
\begin{align} \label{light-phi}
& \mathcal{I}_{\varphi}(\nu_1,\nu_2,0,\mu) = \mathcal{I}_0 \left(\nu_1 , \nu_2 , 0 ,\mu \right) 
 \notag\\& 
-\frac{\alpha _s C_F}{8 \pi } \frac{1}{\epsilon_{\text{IR}}}
\left\{
2 \int_0^1 d \eta  \left(\frac{1-\eta }{\eta }\right)_+ 
\left \{  \left(\mathcal{I}_0 \left((1-\eta ) \nu_1+\eta  \nu_2 , \nu_2 , 0 , \mu \right)+
\mathcal{I}_0 \left(\nu_1 , (1-\eta ) \nu_2+\eta  \nu_1 , 0 , \mu \right) \right) \right.\right.
\notag\\& 
\left.\left.
 + 
 \left(\mathcal{I}_0 \left((1-\eta ) \nu_1 , \nu_2 , 0 , \mu \right)+\mathcal{I}_0 \left(\nu_1 , \nu_2 , \eta  \nu_1 , \mu \right)\right)
+
\left(
\mathcal{I}_0 \left(\nu_1 , (1-\eta ) \nu_2 , 0 , \mu 
\right)
+\mathcal{I}_0 \left(\nu_1 , \nu_2 , \eta  \nu_2 , \mu \right)
\right)
\right \} \right \} ,
\end{align} 
where $\nu_1\equiv z_1 n\cdot P$, $\nu_2\equiv z_2 n\cdot P$, and $ {\mathcal{I}}_0$ stands for tree-level matrix element. It is evident that the lightcone correlator shares the same IR poles in the form of
$\displaystyle \frac{1}{\epsilon_{\rm IR}}$ as the spatial correlator.
The lightcone correlator can be normalized by dividing it by its zero-momentum matrix element: 
  \begin{equation}
 \mathfrak{I}_{V,A,T,\varphi}(\nu_1,\nu_2,\nu_3,\mu) \equiv \frac{\mathcal{I}_{V,A,T,\varphi}(\nu_1,\nu_2,\nu_3,\mu)}{\mathcal{I}_{V,A,T,\varphi}(0,0,0,\mu)}.     
  \end{equation}
The LCDAs with Fock states in momentum space are defined by
\begin{equation}\label{Fock-LC}
\begin{aligned}
& \Psi_{\overline{\rm{MS}}}^{V,A,T,\varphi}\left(x_1, x_2,x_3=1-x_1-x_2,\mu \right)=
\int_{-\infty}^{+\infty} \frac{n\cdot P d \,  z_1}{2 \pi}  \frac{n\cdot P d \, z_2}{2 \pi}
\times  e^{i x_1 n\cdot P z_1+i x_2 n\cdot P z_2} 
\times \mathfrak{I}_{V,A,T,\varphi}\left(z_1 n\cdot P, z_2 n\cdot P,0,\mu\right).
\end{aligned}
\end{equation}
In the following discussion, the variable $x_3=1-x_1-x_2$ will not be written out explicitly.

\subsubsection{One-loop perturbative results of quasi-DAs under the hybrid renormalization scheme}
Under the hybrid renormalization scheme, the normalized $\overline{\rm{MS}}$ renormalized spatial correlators $\widehat{M}$ should be further manipulated by dividing it by its zero-momentum matrix element:
\begin{equation}
    \mathfrak{M}_{V,A,T,\varphi}(z_1,z_2,0,P^z,\mu)=\displaystyle\frac{\widehat{M}_{V,A,T,\varphi}(z_1,z_2,0,P^z,\mu)}{\widehat{M}_{V,A,T,\varphi}(z_1,z_2,0,0,\mu)}.
\end{equation}
Now the perturbative results of spatial correlators under the hybrid renormalization scheme can be expressed as
 \begin{align} \label{Fock-hybrid-VA}
 \mathfrak{M}_{V(A)}&(z_1,z_2,0,P^z,\mu)= 
 \mathfrak{M}_0 \left(z_1,z_2,0,P^z,\mu\right) 
 \notag\\ 
-&\frac{\alpha _s C_F}{8 \pi } \int_0^1 d \eta _1 \int_0^{1-\eta _1} d \eta _2   
 \notag\\ 
\times &
\left\{
\left (L_1^\text{IR}-1+\frac{1}{\epsilon_{\mathrm{IR}}}\right )
\left(
\mathfrak{M}_0 \left(\left(1-\eta _1\right) z_1 , z_2 , \eta _2 z_1, 0,P^2,\mu \right) 
-
\mathfrak{M}_0 \left(z_1 , z_2 , 0,P^2,\mu\right)
\right)
\right.
\notag\\&
+
\left (L_2^\text{IR}-1+\frac{1}{\epsilon_{\mathrm{IR}}}\right )
\left(
\mathfrak{M}_0 \left(z_1 , \left(1-\eta _1\right) z_2 , \eta _2 z_2 ,P^2,\mu\right)
-
\mathfrak{M}_0 \left(z_1 , z_2 , 0,P^2,\mu\right)
\right)
\notag\\&
\left.
+2 
\left (L_{12}^\text{IR}-3+\frac{1}{\epsilon_{\mathrm{IR}}}\right )
\left(
\mathfrak{M}_0 \left(\left(1-\eta _1\right) z_1+\eta _1 z_2 , \left(1-\eta _2\right) z_2+\eta _2 z_1 , 0 ,P^2,\mu\right) 
-
\mathfrak{M}_0 \left( z_1,z_2 , 0 ,P^2,\mu \right)
\right)
\right\}
 \notag\\ 
 -&\frac{\alpha _s C_F}{4 \pi } \int_0^1 d \eta 
 \notag\\
 \times &\left\{
 \left(
 \mathfrak{M}_0 \left(z_1 , (1-\eta ) z_2 , 0 ,P^2,\mu \right)
 +
 \mathfrak{M}_0 \left(z_1 , z_2 , \eta  z_2 ,P^2,\mu \right)
\right)
\left\{
\left( L_2^\text{IR}+1+\frac{1}{\epsilon_{\mathrm{IR}}}
\right)
\left(\frac{1-\eta }{\eta }\right)_+ 
+2 \left(\frac{\ln  \eta  }{\eta }\right)_+ \right\}  
\right.
\notag\\& 
+\left(
\mathfrak{M}_0 \left((1-\eta ) z_1 , z_2 , 0 ,P^2,\mu \right)
+\mathfrak{M}_0 \left(z_1 , z_2 , \eta  z_1 ,P^2,\mu \right) 
\right)
\left\{ 
\left (L_1^\text{IR}+1+\frac{1}{\epsilon_{\mathrm{IR}}}\right )
\left(\frac{1-\eta }{\eta }\right)_+
+2 \left(\frac{\ln  \eta }{\eta }\right)_+\right\} 
 \notag\\ &
\left.
+\left(
\mathfrak{M}_0 \left((1-\eta ) z_1+\eta z_2 , z_2 , 0 ,P^2,\mu \right)
+
\mathfrak{M}_0 \left(z_1 , (1-\eta ) z_2+\eta z_1 , 0 ,P^2,\mu \right)
\right)
\left\{
\left (L_{12}^\text{IR}+1+\frac{1}{\epsilon_{\mathrm{IR}}}\right )
\left(\frac{1-\eta }{\eta }\right)_+ 
+2 \left(\frac{\ln  \eta }{\eta }\right)_+ \right\}
\right\},
\end{align}    
 \begin{align} \label{Fock-hybrid-T}
 \mathfrak{M}_T&(z_1,z_2,0,P^z,\mu)=
 \mathfrak{M}_0 \left(z_1,z_2,0,P^z,\mu\right) 
 \notag\\ 
-&\frac{\alpha _s C_F}{8 \pi } \int_0^1 d \eta _1 \int_0^{1-\eta _1} d \eta _2   
 \notag\\ 
\times &
\left\{
\left (L_1^\text{IR}-1+\frac{1}{\epsilon_{\mathrm{IR}}}\right )
\left(
\mathfrak{M}_0 \left(\left(1-\eta _1\right) z_1 , z_2 , \eta _2 z_1, 0,P^2,\mu \right) 
-
\mathfrak{M}_0 \left(z_1 , z_2 , 0,P^2,\mu\right)
\right)
\right.
\notag\\&
\left.+
\left (L_2^\text{IR}-1+\frac{1}{\epsilon_{\mathrm{IR}}}\right )
\left(
\mathfrak{M}_0 \left(z_1 , \left(1-\eta _1\right) z_2 , \eta _2 z_2 ,P^2,\mu\right)
-
\mathfrak{M}_0 \left(z_1 , z_2 , 0,P^2,\mu\right)
\right)
\right\}
 \notag\\ 
 -&\frac{\alpha _s C_F}{4 \pi } \int_0^1 d \eta 
 \notag\\
 \times &\left\{
 \left(
 \mathfrak{M}_0 \left(z_1 , (1-\eta ) z_2 , 0 ,P^2,\mu \right)
 +
 \mathfrak{M}_0 \left(z_1 , z_2 , \eta  z_2 ,P^2,\mu \right)
\right)
\left\{
\left( L_2^\text{IR}+1+\frac{1}{\epsilon_{\mathrm{IR}}}
\right)
\left(\frac{1-\eta }{\eta }\right)_+ 
+2 \left(\frac{\ln  \eta  }{\eta }\right)_+ \right\}  
\right.
\notag\\& 
+\left(
\mathfrak{M}_0 \left((1-\eta ) z_1 , z_2 , 0 ,P^2,\mu \right)
+\mathfrak{M}_0 \left(z_1 , z_2 , \eta  z_1 ,P^2,\mu \right) 
\right)
\left\{ 
\left (L_1^\text{IR}+1+\frac{1}{\epsilon_{\mathrm{IR}}}\right )
\left(\frac{1-\eta }{\eta }\right)_+
+2 \left(\frac{\ln  \eta }{\eta }\right)_+\right\} 
 \notag\\ &
\left.
+\left(
\mathfrak{M}_0 \left((1-\eta ) z_1+\eta z_2 , z_2 , 0 ,P^2,\mu \right)
+
\mathfrak{M}_0 \left(z_1 , (1-\eta ) z_2+\eta z_1 , 0 ,P^2,\mu \right)
\right)
\left\{
\left (L_{12}^\text{IR}+1+\frac{1}{\epsilon_{\mathrm{IR}}}\right )
\left(\frac{1-\eta }{\eta }\right)_+ 
+2 \left(\frac{\ln  \eta }{\eta }\right)_+ \right\}
\right\},
\end{align}       
\begin{align} \label{Fock-hybrid-phi}
 \mathfrak{M}_{\varphi}&(z_1,z_2,0,P^z,\mu)=
 \mathfrak{M}_0 \left(z_1,z_2,0,P^z,\mu\right) 
 \notag\\ 
 -&\frac{\alpha _s C_F}{4 \pi } \int_0^1 d \eta 
 \notag\\
 \times &\left\{
 \left(
 \mathfrak{M}_0 \left(z_1 , (1-\eta ) z_2 , 0 ,P^2,\mu \right)
 +
 \mathfrak{M}_0 \left(z_1 , z_2 , \eta  z_2 ,P^2,\mu \right)
\right)
\left\{
\left( L_2^\text{IR}+1+\frac{1}{\epsilon_{\mathrm{IR}}}
\right)
\left(\frac{1-\eta }{\eta }\right)_+ 
+2 \left(\frac{\ln  \eta  }{\eta }\right)_+ \right\}  
\right.
\notag\\& 
+\left(
\mathfrak{M}_0 \left((1-\eta ) z_1 , z_2 , 0 ,P^2,\mu \right)
+\mathfrak{M}_0 \left(z_1 , z_2 , \eta  z_1 ,P^2,\mu \right) 
\right)
\left\{ 
\left (L_1^\text{IR}+1+\frac{1}{\epsilon_{\mathrm{IR}}}\right )
\left(\frac{1-\eta }{\eta }\right)_+
+2 \left(\frac{\ln  \eta }{\eta }\right)_+\right\} 
 \notag\\ &
\left.
+\left(
\mathfrak{M}_0 \left((1-\eta ) z_1+\eta z_2 , z_2 , 0 ,P^2,\mu \right)
+
\mathfrak{M}_0 \left(z_1 , (1-\eta ) z_2+\eta z_1 , 0 ,P^2,\mu \right)
\right)
\left\{
\left (L_{12}^\text{IR}+1+\frac{1}{\epsilon_{\mathrm{IR}}}\right )
\left(\frac{1-\eta }{\eta }\right)_+ 
+2 \left(\frac{\ln  \eta }{\eta }\right)_+ \right\}
\right\}.
\end{align}       
The corresponding quasi-DAs in momentum space are defined as
\begin{equation}\label{Fock-quasi}
\begin{aligned}
& \widetilde \Psi^{V,A,T,\varphi}_H\left(x_1, x_2,x_3=1-x_1-x_2,P^z,\mu \right)=
\int_{-\infty}^{+\infty} \frac{P^zd \,  z_1}{2 \pi}  \frac{P^zd \, z_2}{2 \pi}
\times  e^{-i x_1 P^z z_1-i x_2 P^z z_2} 
\times {\mathfrak{M}}_{V,A,T,\varphi}\left(z_1, z_2,0,P^z,\mu\right).
\end{aligned}
\end{equation}

\subsubsection{Hard kernels}
Given the perturbative LCDAs $\Psi^{V,A,T,\varphi}_{\overline{\rm MS}}$ with Fock states  renormalized under the ${\overline{\rm MS}}$ scheme and the perturbative quasi-DAs $\widetilde\Psi_H^{V,A,T,\varphi}$ renormalized under the hybrid renormalization scheme, the corresponding hard kernels up to one-loop accuracy can then be extracted through Eq.~(\ref{Fock-mai}) as
\begin{align}
\mathcal{C}_{V,A}\left(x_1, x_2, y_1, y_2, P^z,\mu\right) & = \delta\left(x_1-y_1\right) \delta\left(x_2-y_2\right)
+\frac{\alpha_s C_F}{4 \pi}C_{1 V,A}\left(x_1, x_2, y_1, y_2, P^z,\mu\right)
\notag\\&
+\frac{\alpha_s C_F}{4 \pi} 
 \times\left[C_2\left(x_1, x_2, y_1, y_2,P^z,\mu\right) \delta \left(x_2-y_2\right)\right.
 \notag\\&
 \left.+C_3\left(x_1, x_2, y_1, y_2,P^z,\mu\right) \delta\left(x_3-y_3\right) 
 +\left\{x_1 \leftrightarrow x_2, y_1 \leftrightarrow y_2\right\}\right]_{\oplus},\label{ker-va}
 \\
 \mathcal{C}_{T}\left(x_1, x_2, y_1, y_2, P^z,\mu\right) 
 & = \delta\left(x_1-y_1\right) \delta\left(x_2-y_2\right)
+\frac{\alpha_s C_F}{4 \pi}C_{1 T}\left(x_1, x_2, y_1, y_2, P^z,\mu\right)
\notag\\&
+\frac{\alpha_s C_F}{4 \pi} 
 \times\left[C_2\left(x_1, x_2, y_1, y_2,P^z,\mu\right) \delta \left(x_2-y_2\right)\right.
 \notag\\&
 \left.+(C_3-C_5)\left(x_1, x_2, y_1, y_2,P^z,\mu\right) \delta\left(x_3-y_3\right) 
 +\left\{x_1 \leftrightarrow x_2, y_1 \leftrightarrow y_2\right\}\right]_{\oplus}, \label{ker-t}
 \\
 \mathcal{C}_{\varphi}\left(x_1, x_2, y_1, y_2, P^z,\mu\right) & = \delta\left(x_1-y_1\right) \delta\left(x_2-y_2\right)
+\frac{\alpha_s C_F}{4 \pi}C_{1 \varphi}\left(x_1, x_2, y_1, y_2, P^z,\mu\right)
\notag\\&
+\frac{\alpha_s C_F}{4 \pi} 
 \times\left[(C_2-C_4)\left(x_1, x_2, y_1, y_2,P^z,\mu\right) \delta \left(x_2-y_2\right)\right.
 \notag\\&
 \left.+(C_3-C_5)\left(x_1, x_2, y_1, y_2,P^z,\mu\right) \delta\left(x_3-y_3\right) 
 +\left\{x_1 \leftrightarrow x_2, y_1 \leftrightarrow y_2\right\}\right]_{\oplus},\label{ker-phi}
\end{align}
where $C_{1V,A}$, $C_{1T}$ and $C_{1 \varphi}$ are
\begin{align}\label{eq}
&\mathcal{C}_{1 V,A}(x_1,x_2,y_1,y_2,P^z,\mu) =2 (P^z)^2 \Bigg[ I^{V/A}_{\rm H} [(x_1-y_1)P^z,(x_2-y_2)P^z] + I^{V/A}_{\rm HSI} [(x_1-y_1)P^z,(x_2-y_2)P^z] \notag\\
&+ I^{V/A}_{\rm HSII} [(x_1-y_1)P^z,(x_2-y_2)P^z] + I^{V/A}_{\rm HSIII} [(x_1-y_1)P^z,(x_2-y_2)P^z] + I^{V/A}_{\rm HSIV} [(x_1-y_1)P^z,(x_2-y_2)P^z] \notag\\
&+ I^{V/A}_{\rm S} [(x_1-y_1)P^z,(x_2-y_2)P^z] + \delta[(x_1-y_1)P^z] \delta[(x_2-y_2)P^z] \left(\frac{5}{2} \ln\left(\frac{\mu^2 e^{2 \gamma_E}}{4}\right) + 4\right) \Bigg],
\\
&\mathcal{C}_{1 T}(x_1,x_2,y_1,y_2,P^z,\mu) = 2 (P^z)^2  \Bigg[ I^{T}_{\rm H} [(x_1-y_1)P^z,(x_2-y_2)P^z] + I^{T}_{\rm HSI} [(x_1-y_1)P^z,(x_2-y_2)P^z] \notag\\
&+ I^{T}_{\rm HSII} [(x_1-y_1)P^z,(x_2-y_2)P^z] + I^{T}_{\rm HSIII} [(x_1-y_1)P^z,(x_2-y_2)P^z] + I^{T}_{\rm HSIV} [(x_1-y_1)P^z,(x_2-y_2)P^z] \notag\\
&+ I^{T}_{\rm S} [(x_1-y_1)P^z,(x_2-y_2)P^z] + \delta[(x_1-y_1)P^z] \delta[(x_2-y_2)P^z] \left(\frac{9}{4} \ln\left(\frac{\mu^2 e^{2 \gamma_E}}{4}\right) + \frac{13}{4}\right) \Bigg],
\\
&\mathcal{C}_{1 \varphi}(x_1,x_2,y_1,y_2,P^z,\mu) = 2 (P^z)^2  \Bigg[ I^{\varphi}_{\rm H} [(x_1-y_1)P^z,(x_2-y_2)P^z] + I^{\varphi}_{\rm HSI} [(x_1-y_1)P^z,(x_2-y_2)P^z] \notag\\
&+ I^{\varphi}_{\rm HSII} [(x_1-y_1)P^z,(x_2-y_2)P^z] + I^{\varphi}_{\rm HSIII} [(x_1-y_1)P^z,(x_2-y_2)P^z] + I^{\varphi}_{\rm HSIV} [(x_1-y_1)P^z,(x_2-y_2)P^z] \notag\\
&+ I^{\varphi}_{\rm S} [(x_1-y_1)P^z,(x_2-y_2)P^z] + \delta[(x_1-y_1)P^z] \delta[(x_2-y_2)P^z] \left(\frac{9}{4} \ln\left(\frac{\mu^2 e^{2 \gamma_E}}{4}\right) + 3\right) \Bigg],
\end{align}
and $\oplus$ denotes a double plus function   defined as 
\begin{equation} \label{dpf}
{\left[g\left(x_1, x_2, y_1, y_2\right)\right]_{\oplus}=}   g\left(x_1, x_2, y_1, y_2\right) 
 -\delta\left(x_1-y_1\right) \delta\left(x_2-y_2\right) 
 \int d x_1' d x_2' g\left(x_1', x_2', y_1, y_2\right). 
\end{equation} 
The functions $I_{\rm H},I_{\rm HSI},I_{\rm HSII},I_{\rm HSIII},I_{\rm HSIV}$ and $I_{\rm S}$ for each LCDA are collected in Appendix.~\ref{sec:appe kernel}.
A crucial requirement for a well-defined hard kernel is the absence of linearly decaying terms, such as \(\sim \frac{1}{|x'_1|}\) and \(\frac{1}{|x'_2|}\) for \(|x'_1| \gg 1\) and \(|x'_2| \gg 1\), respectively. 
It can be verified that these linearly decaying terms vanish in these kernel results. 
This cancellation is equivalent to the elimination of logarithmic terms like \(\ln(\mu^2z_1^2)\), \(\ln(\mu^2z_2^2)\), and \(\ln(\mu^2(z_1-z_2)^2)\) through the ratio scheme when $z_1\to 0,z_2\to 0$ and $(z_1-z_2)\to 0$.

\section{summary}
In this study, we focus on the extraction of the hard kernels of light octet and decuplet baryons to the next-to-leading order in $\alpha_s$, facilitating the simulation of LCDAs via Lattice QCD. We commence by offering comprehensive definitions of LCDAs, elucidating their symmetry properties and asymptotic behaviors. Subsequently, we have defined their corresponding spatial correlators, followed by the computation of all typical spatial correlators crucial for constructing the lowest-lying baryons up to the next-to-leading order. These correlators are renormalized in the hybrid renormalization scheme to effectively remove the singular UV logarithms. The hard kernels of light octet and decuplet baryons are then derived through one-loop matching between LCDAs and corresponding quasi-DAs with lowest-order Fock states.

Utilizing these hard kernels, we can then determine the LCDAs of the lowest-lying octet and decuplet baryons from lattice-simulated spatial correlators. These outcomes not only serve as robust inputs for advancing studies on light baryons in Lattice QCD but also contribute to the understanding of model construction phenomena. However, certain issues remain unaddressed in the extraction of baryon LCDAs. Theoretical challenges arise particularly around the matching equation (Eq.~(\ref{matching})) being invalid in the threshold of $y_i\to x_i$, where the physical scales undergo change, necessitating potential modifications to factorization.  Additionally, the validity of an all-order hybrid renormalization scheme needs to be proved. 
In practical terms, while the current hybrid renormalization scheme is theoretically robust, its implementation in Lattice QCD is impeded by the intricate process of dividing regions. Thus, a more straightforward and implementable renormalization scheme is desired.

In our subsequent work, we intend to study the extraction of baryon LCDAs near the threshold, and  investigate the moments of octet and decuplet baryon LCDAs. Furthermore, we will explore alternative renormalization schemes that offer greater simplicity and ease of implementation.
\section*{Acknowledgments}

We thank Zhi-Fu Deng  and Yushan Su for the collaboration of Refs.~\cite{Deng:2023csv,Han:2023xbl} and valuable discussions. This work is supported in part by Natural Science Foundation of China under grant No.12125503, 12147140,  12205180 and 12335003. J.Z. is also partially supported by the Project funded by China Postdoctoral Science Foundation under Grant No. 2022M712088.

\appendix

\section{The $\rm{SU(3)}$\label{sec:wf}  flavor and $\rm{SU(2)}$ spin wave functions }\label{sec-quarkmodel}
To construct the light-front wave function of octet and decuplet baryons, we need to know $\rm{SU(3)}$ flavor and $\rm{SU(2)}$ spin wave functions.
The flavor wave functions $|S,B\rangle,|A,B\rangle,|MS,B\rangle $ and $|MA,B\rangle$ of each baryons are defined in TABLE II and TABLE III, and the spin wave functions are defined in TABLE IV.

\begin{table}[!h]
\renewcommand{\arraystretch}{2} 
        \centering
\begin{tabular}{|p{0.48\textwidth}|p{0.52\textwidth}|}
\hline 
 \multicolumn{2}{|l|}{ \ \ \ \ \ \ \ \ \ \ \ \ \ \ \ \ \ \ \ \ \ \ \ \ \ \ \ \ \ \ \ \ \ \ \ \ \ \ \ \ \ \ \ \ \ \ \ \ \ \ \ \ \ \ \ \ \ \ \ \ \ \ \ \ \ \ \ \ \ \ \ \ $\displaystyle |S\rangle $} \\
\hline 
 {\footnotesize $\displaystyle \Delta ^{++} =|uuu\rangle $} & {\footnotesize $\displaystyle \Sigma ^{*-} =\displaystyle\frac{1}{\sqrt{3}}( |dds\rangle +|dsd\rangle +|sdd\rangle )$} \\
\hline 
 {\footnotesize $\displaystyle \Delta ^{+} =\displaystyle\frac{1}{\sqrt{3}}( |uud\rangle +|udu\rangle +|duu\rangle )$} & {\footnotesize $\displaystyle \Xi ^{*0} =\displaystyle\frac{1}{\sqrt{3}}( |ssu\rangle +|sus\rangle +|uss\rangle )$} \\
\hline 
 {\footnotesize $\displaystyle \Delta ^{0} =\displaystyle\frac{1}{\sqrt{3}}( |ddu\rangle +|dud\rangle +|udd\rangle )$} & {\footnotesize $\displaystyle \Xi ^{*-} =\displaystyle\frac{1}{\sqrt{3}}( |ssd\rangle +|sds\rangle +|dss\rangle )$} \\
\hline 
 {\footnotesize $\displaystyle \Delta ^{-} =|ddd\rangle $} & {\footnotesize $\displaystyle \Omega ^{*-} =|sss\rangle $} \\
\hline 
{\footnotesize $\displaystyle \Sigma ^{*+} =\displaystyle\frac{1}{\sqrt{3}}( |uus\rangle +|usu\rangle +|suu\rangle )$} & {\footnotesize $\displaystyle \Sigma ^{*0} =\displaystyle\frac{1}{\sqrt{6}}( |uds\rangle +|usd\rangle +|dsu\rangle +|dus\rangle +|sud\rangle +|sdu\rangle )$} \\
\hline 
 \multicolumn{2}{|l|}
 {
 \ \ \ \ \ \ \ \ \ \ \ \ \  \ \ \ \ \ \ \ \ \ \ \ \ \ \ \ \ \ \ \ \ \ \ \ \
$\displaystyle |A\rangle =\frac{1}{\sqrt{6}}( |uds\rangle -|usd\rangle +|dsu\rangle -|dus\rangle +|sud\rangle -|sdu\rangle $

} 
\\
\hline
\end{tabular}
        \caption{Flavor wave functions of $|S\rangle$ and $|A\rangle$}
        \end{table}

{\small 
\begin{table}[!h]
\renewcommand{\arraystretch}{2}
        \centering     
\begin{tabular}{|p{0.50\textwidth}|p{0.50\textwidth}|}
\hline 
 {\footnotesize $\displaystyle |MS\rangle $} & {\footnotesize $\displaystyle |MA\rangle $} \\
\hline 
 {\footnotesize $\displaystyle p=\frac{1}{\sqrt{6}}( 2|uud\rangle -|udu\rangle -|duu\rangle )$} & {\footnotesize $\displaystyle p=\frac{1}{\sqrt{2}}( |udu\rangle -|duu\rangle )$} \\
\hline 
 {\footnotesize $\displaystyle n=\frac{1}{\sqrt{6}}( -2|ddu\rangle +|dud\rangle +|udd\rangle )$} & {\footnotesize $\displaystyle n=\frac{1}{\sqrt{2}}( |udd\rangle -|dud\rangle )$} \\
\hline 
 {\footnotesize $\displaystyle \Sigma ^{+} =\frac{1}{\sqrt{6}}( 2|uus\rangle -|usu\rangle -|suu\rangle )$} & {\footnotesize $\displaystyle \Sigma ^{+} =\frac{1}{\sqrt{2}}( |usu\rangle -|suu\rangle )$} \\
\hline 
 {\footnotesize $\displaystyle \Sigma ^{0} =\frac{1}{\sqrt{12}}( 2|uds\rangle -|usd\rangle -|dsu\rangle +2|dus\rangle -|sud\rangle -|sdu\rangle )$} & {\footnotesize $\displaystyle \Sigma ^{0} =\frac{1}{2}( |usd\rangle +|dsu\rangle -|sdu\rangle -|sud\rangle )$} \\
\hline 
 {\footnotesize $\displaystyle \Sigma ^{-} =\frac{1}{\sqrt{6}}( 2|dds\rangle -|dsd\rangle -|sdd\rangle )$} & {\footnotesize $\displaystyle \Sigma ^{-} =\frac{1}{\sqrt{2}}( |dsd\rangle -|sdd\rangle )$} \\
\hline 
 {\footnotesize $\displaystyle \Lambda ^{0} =\frac{1}{2}( |usd\rangle +|sud\rangle -|sdu\rangle -|dsu\rangle )$} & {\footnotesize $\displaystyle \Lambda ^{0} =\frac{1}{\sqrt{12}}( 2|uds\rangle -|dsu\rangle -|sud\rangle -2|dus\rangle +|sdu\rangle +|usd\rangle )$} \\
\hline 
 {\footnotesize $\displaystyle \Xi ^{0} =\frac{1}{\sqrt{6}}( |sus\rangle +|uss\rangle -2|ssu\rangle )$} & {\footnotesize $\displaystyle \Xi ^{0} =\frac{1}{\sqrt{2}}( |uss\rangle -|sus\rangle )$} \\
\hline 
 {\footnotesize $\displaystyle \Xi ^{-} =\frac{1}{\sqrt{6}}( |sds\rangle +|dss\rangle -2|ssd\rangle )$} & {\footnotesize $\displaystyle \Xi ^{-} =\frac{1}{\sqrt{2}}( |dss\rangle -|sds\rangle )$} \\
 \hline
\end{tabular}
        \caption{Flavor wave functions of $|MS\rangle$ and $|MA\rangle$}
        \end{table}
        }
It is evident that the state $|S\rangle$ exhibits total symmetry in three valence quarks, while $|A\rangle$ is total anti-symmetry. Moreover, $|MS\rangle$($|MA\rangle$) display (anti-)symmetry concerning the first two quarks. This observation imposes constraints on the symmetry properties of the $V$, $A$, $T$ and $\varphi$.

{\small \begin{table}[!h]
        \centering
        
\begin{tabular}{|p{0.50\textwidth}|p{0.50\textwidth}|}
\hline 
 $\displaystyle 4_{S} :$$\displaystyle |3/2,3/2\rangle $ & $\displaystyle |\uparrow \uparrow \uparrow \rangle $ \\
\hline 
 $\displaystyle 4_{S} :$$\displaystyle |3/2,1/2\rangle $ & $\displaystyle \frac{1}{\sqrt{3}}( |\uparrow \uparrow \downarrow \rangle +|\uparrow \downarrow \uparrow \rangle +|\downarrow \uparrow \uparrow \rangle $ \\
\hline 
 $\displaystyle 4_{S} :|3/2,-1/2\rangle $ & $\displaystyle \frac{1}{\sqrt{3}}( |\uparrow \downarrow \downarrow \rangle +|\downarrow \downarrow \uparrow \rangle +|\downarrow \downarrow \uparrow \rangle $ \\
\hline 
 $\displaystyle 4_{S} :|3/2,-3/2\rangle $ & $\displaystyle |\downarrow \downarrow \downarrow \rangle $ \\
\hline 
 $\displaystyle 2_{MS} :|1/2,1/2\rangle $ & $\displaystyle \frac{1}{\sqrt{6}}( 2|\uparrow \uparrow \downarrow \rangle -|\uparrow \downarrow \uparrow \rangle -|\downarrow \uparrow \uparrow \rangle )$ \\
\hline 
 $\displaystyle 2_{MS} :|1/2,-1/2\rangle $ & $\displaystyle \frac{1}{\sqrt{6}}( -2|\downarrow \downarrow \uparrow \rangle +|\uparrow \downarrow \downarrow \rangle +|\downarrow \uparrow \downarrow \rangle )$ \\
\hline 
 $\displaystyle 2_{MA} :|1/2,1/2\rangle $ & $\displaystyle \frac{1}{\sqrt{2}}( |\uparrow \downarrow \uparrow \rangle -|\downarrow \uparrow \uparrow \rangle )$ \\
\hline 
 $\displaystyle 2_{MA} :|1/2,-1/2\rangle $ & $\displaystyle \frac{1}{\sqrt{2}}( |\uparrow \downarrow \downarrow \rangle -|\downarrow \uparrow \downarrow \rangle )$ \\
 \hline
\end{tabular}
\caption{Spin wave functions of $\displaystyle 4_{S}$, $\displaystyle 2_{MS}$ and $\displaystyle 2_{MA}$}
        \end{table}}

\clearpage

\section{Hard kernels}\label{sec:appe kernel}

In the hybrid renormalization scheme, the counterterm in the hard kernel can be split into different regions.
In this appendix, we provide the explicit expressions for these terms of each quasi-DAs. 

\subsection{Building blocks of the hard kernel calculation}
To calculated the hard kernels in Eqs.~(\ref{ker-va}, \ref{ker-t}, \ref{ker-phi}), we need to know the explicitly formula for the $C_{2-5}$ and various integrate therein.
The $C_{2-5}$ can be obtained from the hard kernel in the $\overline{\text{MS}}$ scheme.
In this scheme, the  one-loop hard kernel has been obtained~\cite{Deng:2023csv}.
The $C_{2-5}$ are 
\begin{align}
& C_2\left(x_1, x_2, y_1, y_2,P^z,\mu\right)= \notag\\
& \left\{\begin{array}{l}
\displaystyle\frac{\left(x_1+y_1\right)\left(x_3+y_3\right) \ln \frac{y_1-x_1}{-x_1}}{y_1\left(y_1-x_1\right) y_3}-\frac{x_3\left(x_1+y_1+2 y_3\right) \ln \frac{x_3}{-x_1}}{\left(y_1-x_1\right) y_3\left(y_1+y_3\right)}, x_1<0 
\\
\displaystyle\frac{\left(x_1-3 y_1-2 y_3\right) x_1}{y_1\left(x_3-y_3\right)\left(y_1+y_3\right)}-\frac{\left[\left(x_3-y_3\right)^2-2 x_3 y_1\right] \ln \frac{x_3-y_3}{x_3}}{y_1\left(x_3-y_3\right) y_3}+\frac{2 x_1 \ln \frac{4 x_1\left(x_3-y_3\right) P_z^2}{\mu^2}}{y_1\left(x_3-y_3\right)}+\frac{x_1 \ln \frac{4 x_1 x_3 P_z^2}{\mu^2}}{y_1\left(y_1+y_3\right)}, 0<x_1<y_1 
\\
\displaystyle\frac{\left(x_3-2 y_1-3 y_3\right) x_3}{y_3\left(x_1-y_1\right)\left(y_1+y_3\right)}-\frac{\left[\left(x_1-y_1\right)^2-2 x_1 y_3\right] \ln \frac{x_1-y_1}{x_1}}{\left(x_1-y_1\right) y_1 y_3}+\frac{2 x_3 \ln \frac{4 x_3\left(x_1-y_1\right) P_z^2}{\mu^2}}{\left(x_1-y_1\right) y_3}+\frac{x_3 \ln \frac{4 x_1 x_3 P_z^2}{\mu^2}}{y_3\left(y_1+y_3\right)}, y_1<x_1<y_1+y_3 
\\
\displaystyle\frac{\left(x_1+y_1\right)\left(x_3+y_3\right) \ln \frac{y_3-x_3}{-x_3}}{y_1 y_3\left(y_3-x_3\right)}-\frac{x_1\left(x_3+2 y_1+y_3\right) \ln \frac{x_1}{-x_3}}{y_1\left(y_3-x_3\right)\left(y_1+y_3\right)}, x_1>y_1+y_3,
\end{array}\right. 
\end{align}
\begin{align}
& C_3\left(x_1, x_2, y_1, y_2,P^z,\mu\right)= \notag\\
& \left\{\begin{array}{l}
\displaystyle\frac{\left(x_1 x_2+y_1 y_2\right) \ln \frac{x_2-y_2}{x_2}}{y_1\left(x_2-y_2\right) y_2}-\frac{x_1\left(x_2+y_1\right) \ln \frac{-x_1}{x_2}}{y_1\left(x_2-y_2\right)\left(y_1+y_2\right)}, x_1<0 
\\
\displaystyle\frac{1}{x_1-y_1}+\frac{2 x_1+x_2}{y_1\left(y_1+y_2\right)}+\frac{\left[\left(x_1+y_2\right) y_1-x_1^2\right] \ln \frac{x_2-y_2}{x_2}}{y_1\left(x_2-y_2\right) y_2}+\frac{x_1 \ln \frac{4 x_1\left(x_2-y_2\right) P_z^2}{\mu^2}}{y_1\left(x_2-y_2\right)}+\frac{x_1 \ln \frac{4 x_1 x_2 P_z^2}{\mu^2}}{y_1\left(y_1+y_2\right)}, 0<x_1<y_1 
\\
\displaystyle\frac{1}{x_2-y_2}+\frac{x_1+2 x_2}{y_2\left(y_1+y_2\right)}+\frac{\left[\left(x_2+y_1\right) y_2-x_2^2\right] \ln \frac{x_1-y_1}{x_1}}{\left(x_1-y_1\right) y_1 y_2}+\frac{x_2 \ln \frac{4 x_2\left(x_1-y_1\right) P_z^2}{\mu^2}}{\left(x_1-y_1\right) y_2}+\frac{x_2 \ln \frac{4 x_1 x_2 P_z^2}{\mu^2}}{y_2\left(y_1+y_2\right)}, y_1<x_1<y_1+y_2 
\\
\displaystyle\frac{\left(x_1 x_2+y_1 y_2\right) \ln \frac{x_1-y_1}{x_1}}{y_1\left(x_1-y_1\right) y_2}-\frac{x_2\left(x_1+y_2\right) \ln \frac{-x_2}{x_1}}{y_2\left(x_1-y_1\right)\left(y_1+y_2\right)}, x_1>y_1+y_2
,
\end{array}\right. 
\end{align}

\begin{align}
& C_4\left(x_1, x_2, y_1, y_2,P^z,\mu\right)= 
& \notag\\
& \left\{\begin{array}{l}
\displaystyle{\left[\frac{x_1 \ln \frac{-x_1}{x_3}}{\left(1-y_2\right) y_3}-\frac{x_1 \ln \frac{-x_1}{y_{1}-x_1}}{y_{1} y_{3}}-\frac{\ln \frac{y_{1}-x_1}{x_3}}{y_3}\right], \quad x_1<0} 
\\
\displaystyle{\left[
\frac{x_1\left(\ln \frac{\left(y_1-x_1\right) x_1}{\mu^2 /\left(2 P^z\right)^2}-1\right)}{y_1 y_3}-\frac{x_1\left(\ln \frac{x_1 x_3}{\mu^2 /\left(2 P^z\right)^2}-1\right)}{\left(1-y_2\right) y_3}-\frac{\ln \frac{y_1-x_1}{x_3}}{y_3}
\right], 0<x_1<y_1} 
\\
\displaystyle{\left[\frac{x_1 \ln \frac{x_1}{x_1-y_1}}{y_1 y_3}-\frac{x_1\left(\ln \frac{x_1 x_3}{\mu^2 /\left(2 P^z\right)^2}-1\right)}{\left(1-y_2\right) y_3}+\frac{\ln \frac{\left(x_1-y_1\right) x_3}{\mu^2 /\left(2 P^z\right)^2}-1}{y_3}
\right],y_1<x_1<1-x_2} 
\\
\displaystyle{\left[\frac{x_1 \ln\frac{x_1}{x_1-x_{10}}}{y_1 y_3}+\frac{x_1 \ln \frac{-x_3}{x_1}}{\left(1-y_2\right) y_3}+\frac{\ln \frac{x_1-y_1}{-x_3}}{y_3}\right]  \quad x_1>1-x_2,}
\end{array}\right.
\end{align}
\begin{align}
&C_5\left(x_1, x_2, y_1, y_2,P^z,\mu\right)=\notag
\\&
\left\{\begin{array}{l}
\displaystyle{\left[\frac{x_1}{y_1\left(y_1+y_2\right)} \ln \frac{x_2}{-x_1}+\frac{y_2-x_2}{y_1 y_2} \ln \frac{x_2-y_2}{x_2}\right], \quad x_1<0} 
\\
\displaystyle\frac{x_{1}\left(\ln\left(\frac{x_{2}}{\mu ^{2}/( 2 P_{z})^{2}}\right) +1\right)}{y_{1}( y_{1} +y_{2})} +\frac{x_{2}\ln x_{2}}{y_{2}( y_{1} +y_{2})} -\frac{( x_{2} -y_{2})\ln( x_{2} -y_{2})}{y_{1} y_{2}}, 0<x_1<y_1
\\
\displaystyle\frac{x_{2}\left(\ln\left(\frac{x_{1}}{\mu ^{2}/( 2 P_{z})^{2}}\right) +1\right)}{y_{2}( y_{1} +y_{2})} +\frac{x_{1}\ln x_{1}}{y_{1}( y_{1} +y_{2})} -\frac{( x_{1} -y_{1})\ln( x_{1} -y_{1})}{y_{1} y_{2}}, y_1<x_1<x_1+x_2
\\
\displaystyle{\left[\frac{x_1}{y_1\left(y_1+y_2\right)} \ln \frac{x_1}{-x_2}+\frac{y_2-x_2}{y_1 y_2} \ln \frac{-x_2}{y_2-x_2}\right],\quad x_1>x_1+x_2}.
\end{array}\right.
\end{align}

In the hybrid renormalization scheme, a practical approach for computing the hard kernel involves establishing a set of master integrals. These master integrals are presented as
\begin{align}
\text{I1}\left(\left\{L_2,L_1\right\},p\right) \equiv &\int_{L_1}^{L_2} \frac{d z}{2\pi} e^{-i p z} \ln[z^2] 
\notag\\
=&-\frac{i \left(2 \left(\gamma_E +\log \left(-i L_2 p\right)\right)+\left(-1+e^{i L_2 p}\right) \log \left(L_2^2\right)+2 \Gamma \left(0,-i p L_2\right)\right)}{2 \pi  p}
\notag\\
&+\frac{i \left(2 \left(\gamma_E +\log \left(-i L_1 p\right)\right)+\left(-1+e^{i L_1 p}\right) \log \left(L_1^2\right)+2 \Gamma \left(0,-i p L_1\right)\right)}{2\pi  p},
\end{align}
\begin{equation}
\text{I0}\left(\left\{L_2,L_1\right\},p\right) 
\equiv \int_{L_1}^{L_2} \frac{d z}{2\pi} e^{i p z}
=-\frac{i \left(-1+e^{i L_2 p}\right)}{2 \pi  p}+\frac{i \left(-1+e^{i L_1 p}\right)}{2 \pi  p}.
\end{equation}
\begin{align}
\text{I1t}\left(\left\{L,-L\right\},p\right) 
&\equiv i[\text{I1}\left(\left\{L,0\right\},p\right) -\text{I1}\left(\left\{0,-L\right\},p\right)] 
\notag\\&
= \frac{2 (-\text{Ci}(L p)+\log (L) \cos (L p)+\log (p)+\gamma_E )}{\pi  p},
\end{align}
\begin{equation}
\text{I0}\left(\left\{\infty,L\right\},p\right) 
=\frac{\delta (p)}{2}+\frac{i}{2 \pi  p}+\frac{i \left(-1+e^{i L p}\right)}{2 \pi  p},
\end{equation}
\begin{equation}
\text{I0}\left(\left\{-L,-\infty\right\},p\right) 
=-\frac{i \left(-1+e^{-i L p}\right)}{2 \pi  p}+\frac{\delta (p)}{2}-\frac{i}{2 \pi  p},
\end{equation}
\begin{equation}
\text{I0}\left(\left\{\infty,-\infty\right\},p\right) 
=\delta (p).
\end{equation}
These master integrals serve as foundational elements through which the hard kernel can be formulated and expressed.
It should be noted that in the above formulas, the logarithms, trigonometric functions and exponential functions involving dimensional scales are well-defined, provided that the dimensional units remain consistent for distance and momentum. 
Based upon the above formulas, all the integrals in the following formulas can be written down ($z_s>0$).

\subsection{$V$ and $A$}
The $V$ and $A$ for octet and decuplet baryons share the same short distance structure. In the process of hard kernels, the contribution of the zero-momentum matrix element is segmented into six distinct regions, allowing for individual calculation of each segment. The functions in Eq.~(\ref{ker-va}) are
\begin{align}
&I^{V/A}_{\rm H} [p_1,p_2] \equiv \int \frac{d z_1}{2\pi} \frac{d z_2}{2\pi} e^{-i p_1 z_1 -  i p_2 z_2}  \left[ \frac{7}{8} \ln\left(z_1^2\right) + \frac{7}{8} \ln\left(z_2^2\right) + \frac{3}{4} \ln\left((z_1-z_2)^2\right)\right] \notag\\
&\quad \quad \quad \quad \times \left(\theta(2z_s-|z_1|)\theta(z_s-|z_2|)+\theta(z_s-|z_1|)\theta(|z_2|-z_s)\theta(2z_s-|z_2|)\right) \notag\\
&= \frac{7}{8} \left[\text{I0}\left(\left\{2 z_s,-2 z_s\right\},p_2\right) \text{I1}\left(\left\{z_s,-z_s\right\},p_1\right)+\left(\text{I0}\left(\left\{2 z_s,-2
   z_s\right\},p_1\right)-\text{I0}\left(\left\{z_s,-z_s\right\},p_1\right)\right)
   \text{I1}\left(\left\{z_s,-z_s\right\},p_2\right) \right.\notag\\
   &\left.+\text{I0}\left(\left\{z_s,-z_s\right\},p_2\right) \left(\text{I1}\left(\left\{2 z_s,-2
   z_s\right\},p_1\right)-\text{I1}\left(\left\{z_s,-z_s\right\},p_1\right)\right)+\text{I0}\left(\left\{z_s,-z_s\right\},p_1\right) \text{I1}\left(\left\{2
   z_s,-2 z_s\right\},p_2\right)\right] \notag\\
&+\frac{3}{8 \pi  \left(p_1+p_2\right)}\left[ \sin \left(\left(p_1+p_2\right) z_s\right)
   \left[\text{I1}\left(\left\{z_s,-z_s\right\},p_1\right)+\text{I1}\left(\left\{z_s,-z_s\right\},-p_2\right)\right.\right.\notag\\
   &\left.\left.-\text{I1}\left(\left\{2 z_s,-2
   z_s\right\},p_1\right)-\text{I1}\left(\left\{2 z_s,-2 z_s\right\},-p_2\right)+\text{I1}\left(\left\{3 z_s,-3 z_s\right\},p_1\right)+\text{I1}\left(\left\{3
   z_s,-3 z_s\right\},-p_2\right)\right] \right.\notag\\
   &\left.+\sin \left(2 \left(p_1+p_2\right) z_s\right)
   \left[-\text{I1}\left(\left\{z_s,-z_s\right\},p_1\right)-\text{I1}\left(\left\{z_s,-z_s\right\},-p_2\right) \right.\right.\notag\\
   &\left.\left.+\text{I1}\left(\left\{3 z_s,-3
   z_s\right\},p_1\right)+\text{I1}\left(\left\{3 z_s,-3 z_s\right\},-p_2\right)\right] \right.\notag\\
   &\left.+\cos \left(2 \left(p_1+p_2\right) z_s\right)
   \left[-\text{I1t}\left(\left\{z_s,-z_s\right\},p_1\right)+\text{I1t}\left(\left\{z_s,-z_s\right\},-p_2\right) \right.\right.\notag\\
   &\left.\left.+\text{I1t}\left(\left\{3 z_s,-3
   z_s\right\},p_1\right)-\text{I1t}\left(\left\{3 z_s,-3 z_s\right\},-p_2\right)\right] \right.\notag\\
   &\left.+\cos \left(\left(p_1+p_2\right) z_s\right)
   \left[-\text{I1t}\left(\left\{z_s,-z_s\right\},p_1\right)+\text{I1t}\left(\left\{z_s,-z_s\right\},-p_2\right)-\text{I1t}\left(\left\{2 z_s,-2
   z_s\right\},p_1\right)\right.\right.\notag\\
   &\left.\left.+\text{I1t}\left(\left\{2 z_s,-2 z_s\right\},-p_2\right)+\text{I1t}\left(\left\{3 z_s,-3
   z_s\right\},p_1\right)-\text{I1t}\left(\left\{3 z_s,-3 z_s\right\},-p_2\right)\right] \right],
\end{align}
\begin{align}
&I^{V/A}_{\rm HSI} [p_1,p_2] \equiv \int \frac{d z_1}{2\pi} \frac{d z_2}{2\pi} e^{-i p_1 z_1 -  i p_2 z_2} \theta(z_s-|z_1|)\theta(|z_2|-2z_s)\notag
\\&
\quad \quad \quad \quad \, \,\,\times\left[ \frac{7}{8} \ln\left(z_1^2\right) + \frac{7}{8} \ln\left((2 z_s)^2\right) + \frac{3}{4} \ln\left((z_1-2 z_s {\rm sign}[z_2])^2\right)\right] \notag\\
&= \frac{1}{8} \left[6 e^{2 i p_1 z_s} \text{I1}\left(\left\{-z_s,-3 z_s\right\},p_1\right) \text{I0}\left(\left\{\infty ,2 z_s\right\},p_2\right)+6 e^{-2 i p_1
   z_s} \text{I1}\left(\left\{3 z_s,z_s\right\},p_1\right) \text{I0}\left(\left\{-2 z_s,-\infty \right\},p_2\right)\right.\notag\\
   &\left.+7\left(\text{I0}\left(\{\infty ,-\infty
   \},p_2\right)-\text{I0}\left(\left\{2 z_s,-2 z_s\right\},p_2\right)\right) \left(\log \left(4 z_s^2\right)
   \text{I0}\left(\left\{z_s,-z_s\right\},p_1\right)+\text{I1}\left(\left\{z_s,-z_s\right\},p_1\right)\right)\right],
\end{align}
\begin{align}
&I^{V/A}_{\rm HSII} [p_1,p_2] \equiv \int \frac{d z_1}{2\pi} \frac{d z_2}{2\pi} e^{-i p_1 z_1 - i p_2 z_2} \theta(|z_1|-2z_s)\theta(z_s-|z_2|)\notag
\\&
\quad \quad \quad \quad \quad \times  \left[ \frac{7}{8} \ln\left((2 z_s)^2\right) + \frac{7}{8} \ln\left(z_2^2\right) + \frac{3}{4} \ln\left(({\rm sign}[z_1]2 z_s- z_2)^2\right)\right] \notag\\
&= \frac{1}{8} \left[6 e^{2 i p_2 z_s} \text{I1}\left(\left\{-z_s,-3 z_s\right\},p_2\right) \text{I0}\left(\left\{\infty ,2 z_s\right\},p_1\right)+6 e^{-2 i p_2
   z_s} \text{I1}\left(\left\{3 z_s,z_s\right\},p_2\right) \text{I0}\left(\left\{-2 z_s,-\infty \right\},p_1\right) \right.\notag\\
   &\left.+7\left(\text{I0}\left(\{\infty ,-\infty
   \},p_1\right)-\text{I0}\left(\left\{2 z_s,-2 z_s\right\},p_1\right)\right) \left(\log \left(4 z_s^2\right)
   \text{I0}\left(\left\{z_s,-z_s\right\},p_2\right)+\text{I1}\left(\left\{z_s,-z_s\right\},p_2\right)\right)\right],
\end{align}
\begin{align}
&I^{V/A}_{\rm HSIII} [p_1,p_2] \equiv \int \frac{d z_1}{2\pi} \frac{d z_2}{2\pi} e^{-i p_1 z_1 - i p_2 z_2} \theta(|z_1|-z_s)\theta(|z_2|-z_s)\theta(z_s-|z_1-z_2|)\notag 
\\&
\times \left[ \frac{7}{8} \ln \left(\left(z_s+\left(z_1-z_2\right) \theta \left(z_1-z_2\right)\right){}^2\right)+\frac{7}{8} \ln \left(\left(z_s+\left(z_2-z_1\right)
   \theta \left(z_2-z_1\right)\right){}^2\right) +\frac{3}{4} \ln \left(\left(z_1-z_2\right){}^2\right) \right] 
   \notag\\
&= \frac{1}{8} \text{I0}\left(\{\infty ,-\infty \},p_1+p_2\right) \left[7\log \left(z_s^2\right) \text{I0}\left(\left\{z_s,-z_s\right\},\frac{1}{2}
   \left(p_1-p_2\right)\right)+6\,
   \text{I1}\left(\left\{z_s,-z_s\right\},\frac{1}{2} \left(p_1-p_2\right)\right) \right.
   \notag\\
   &\left. +7e^{\frac{1}{2} i \left(p_1-p_2\right) z_s} \text{I1}\left(\left\{-z_s,-2 z_s\right\},\frac{1}{2} \left(p_1-p_2\right)\right)+7e^{-\frac{1}{2} i \left(p_1-p_2\right) z_s} \text{I1}\left(\left\{2
   z_s,z_s\right\},\frac{1}{2} \left(p_1-p_2\right)\right)\right] 
   \notag\\
   &+\frac{i e^{-i \left(p_1+p_2\right) z_s}}{16 \pi  \left(p_1+p_2\right)} \left[-7\log \left(z_s^2\right) \text{I0}\left(\left\{0,-z_s\right\},p_1\right)+7e^{2 i \left(p_1+p_2\right) z_s} \log
   \left(z_s^2\right) \text{I0}\left(\left\{0,-z_s\right\},-p_2\right) \right.
   \notag\\
   &\left.+7e^{2 i \left(p_1+p_2\right) z_s} \log \left(z_s^2\right)
   \text{I0}\left(\left\{z_s,0\right\},p_1\right)-7\log \left(z_s^2\right) \text{I0}\left(\left\{z_s,0\right\},-p_2\right)-6
   \text{I1}\left(\left\{0,-z_s\right\},p_1\right) \right.
   \notag\\
   &\left.+6 e^{2 i \left(p_1+p_2\right) z_s} \text{I1}\left(\left\{0,-z_s\right\},-p_2\right)-7e^{i p_1 z_s}
   \text{I1}\left(\left\{-z_s,-2 z_s\right\},p_1\right)+7e^{i \left(2 p_1+p_2\right) z_s} \text{I1}\left(\left\{-z_s,-2 z_s\right\},-p_2\right) \right.
   \notag\\
   &\left.+6 e^{2 i\left(p_1+p_2\right) z_s} \text{I1}\left(\left\{z_s,0\right\},p_1\right)-6 \text{I1}\left(\left\{z_s,0\right\},-p_2\right)+7e^{i \left(p_1+2 p_2\right) z_s}
   \text{I1}\left(\left\{2 z_s,z_s\right\},p_1\right)-7e^{i p_2 z_s} \text{I1}\left(\left\{2 z_s,z_s\right\},-p_2\right)\right],
\end{align}
\begin{align}
&I^{V/A}_{\rm HSIV} [p_1,p_2] \equiv \int \frac{d z_1}{2\pi} \frac{d z_2}{2\pi} e^{-i p_1 z_1 -  i p_2 z_2} \theta(|z_1|-z_s)\theta(|z_2|-z_s)\theta(z_s-|z_1+z_2|)\notag 
\\&
\times \left[ \frac{7}{8} \ln \left(\left(z_s+\left(z_1+z_2\right) \theta \left(z_1+z_2\right)\right){}^2\right)+\frac{7}{8} \ln \left(\left(-z_s+\left(z_2+z_1\right)
   \theta \left(-z_2-z_1\right)\right){}^2\right)+\frac{3}{4} \ln \left(\left(2 z_s + |z_1+z_2|\right){}^2\right) \right]  
   \notag\\
   &=\frac{1}{16} \text{I0}\left(\{\infty ,-\infty \},\frac{1}{2} \left(p_1-p_2\right)\right) \left[7\log \left(z_s^2\right)
   \text{I0}\left(\left\{z_s,-z_s\right\},\frac{1}{2} \left(p_1+p_2\right)\right)\right.
   \notag\\
   &\left.+6 e^{i \left(p_1+p_2\right) z_s} \text{I1}\left(\left\{-2 z_s,-3z_s\right\},\frac{1}{2} \left(p_1+p_2\right)\right) +7e^{\frac{1}{2} i \left(p_1+p_2\right) z_s} \text{I1}\left(\left\{-z_s,-2 z_s\right\},\frac{1}{2}
   \left(p_1+p_2\right)\right)\right.
   \notag\\
   &\left.+7e^{-\frac{1}{2} i \left(p_1+p_2\right) z_s} \text{I1}\left(\left\{2 z_s,z_s\right\},\frac{1}{2} \left(p_1+p_2\right)\right)+6
   e^{-i \left(p_1+p_2\right) z_s} \text{I1}\left(\left\{3 z_s,2 z_s\right\},\frac{1}{2} \left(p_1+p_2\right)\right)\right] 
   \notag\\
   &+\frac{i e^{-i \left(p_1+p_2\right) z_s}}{16 \pi 
   \left(p_1-p_2\right)} \left[-7e^{2 i p_2 z_s} \log \left(z_s^2\right) \text{I0}\left(\left\{0,-z_s\right\},p_1\right)+7e^{2 i p_1 z_s} \log
   \left(z_s^2\right) \text{I0}\left(\left\{0,-z_s\right\},p_2\right)\right.
   \notag\\
   &\left.+7e^{2 i p_1 z_s} \log \left(z_s^2\right)
   \text{I0}\left(\left\{z_s,0\right\},p_1\right)-7e^{2 i p_2 z_s} \log \left(z_s^2\right) \text{I0}\left(\left\{z_s,0\right\},p_2\right)-6 e^{2 i\left(p_1+p_2\right) z_s} \text{I1}\left(\left\{-2 z_s,-3 z_s\right\},p_1\right)\right.
   \notag\\
   &\left.+6 e^{2 i \left(p_1+p_2\right) z_s} \text{I1}\left(\left\{-2 z_s,-3z_s\right\},p_2\right)-7e^{i \left(p_1+2 p_2\right) z_s} \text{I1}\left(\left\{-z_s,-2 z_s\right\},p_1\right)+7e^{i \left(2 p_1+p_2\right) z_s}
   \text{I1}\left(\left\{-z_s,-2 z_s\right\},p_2\right)\right.
   \notag\\
   &\left.+7e^{i p_1 z_s} \text{I1}\left(\left\{2 z_s,z_s\right\},p_1\right)-7e^{i p_2 z_s} \text{I1}\left(\left\{2
   z_s,z_s\right\},p_2\right)+6 \text{I1}\left(\left\{3 z_s,2 z_s\right\},p_1\right)-6 \text{I1}\left(\left\{3 z_s,2 z_s\right\},p_2\right)\right],
\end{align}
\begin{align}
&I^{V/A}_{\rm S} [p_1,p_2] \equiv \int \frac{d z_1}{2\pi} \frac{d z_2}{2\pi} e^{-i p_1 z_1 -  i p_2 z_2}  \theta(|z_1|-z_s)\theta(|z_2|-z_s)\theta(|z_1-z_2|-z_s)\theta(|z_1+z_2|-z_s) \notag
\\
&
\quad \quad \quad \quad \times\left[ \frac{7}{8} \ln\left(z_s^2\right) + \frac{7}{8} \ln\left((2z_s)^2\right) + \frac{3}{4} \ln\left(({\rm sign}[z_1]z_s-{\rm sign}[z_2]2z_s)^2\right)\right]
\notag\\
&=\delta \left(p_1-p_2\right) \delta \left(p_1+p_2\right) \left(\frac{7}{4} \log \left(4 z_s^4\right)+\frac{3}{4} \log \left(9 z_s^4\right)\right) 
\notag\\
& -\frac{\delta \left(p_1+p_2\right) \left(6 \log \left(z_s^2\right)+7 \log \left(4 z_s^4\right)\right) \sin \left(\frac{1}{2} \left(p_1-p_2\right) z_s\right)}{4
   \pi  \left(p_1-p_2\right)}
   \notag\\
&-\frac{\delta \left(p_1-p_2\right) \left(6 \log \left(9 z_s^2\right)+7 \log \left(4 z_s^4\right)\right) \sin \left(\frac{1}{2}
   \left(p_1+p_2\right) z_s\right)}{4 \pi  \left(p_1+p_2\right)} 
   \notag\\
& -\frac{\delta \left(p_2\right) \left(20 \log \left(z_s\right)+\log (3456)\right) \sin \left(p_1 z_s\right)}{4 \pi  p_1}-\frac{\delta \left(p_1\right) \left(20
   \log \left(z_s\right)+\log (3456)\right) \sin \left(p_2 z_s\right)}{4 \pi  p_2}
   \notag\\
&-\frac{\left(20 \log \left(z_s\right)+\log (128)\right) \left(p_1 \cos \left(\left(p_1+2 p_2\right) z_s\right)+p_2 \cos \left(\left(2 p_1+p_2\right)
   z_s\right)\right)}{8 \pi ^2 p_1 p_2 \left(p_1+p_2\right)}
   \notag\\
&+\frac{\left(20 \log \left(z_s\right)+\log (93312)\right) \left(p_1 \cos \left(\left(p_1-2 p_2\right) z_s\right)-p_2 \cos \left(2 p_1 z_s-p_2 z_s\right)\right)}{8 \pi ^2 p_1 \left(p_1-p_2\right) p_2}.
\end{align}
\subsection{$T$ and $\varphi$}
Due to the spin structures of $T$ and $\varphi$, certain loop corrections are absent, as illustrated in III.C. However, the logarithmic behavior of $z_i$ is consistent for $T$ and $\varphi$. The functions in Eqs.~(\ref{ker-t}, \ref{ker-phi}) are
\begin{align}
&I^{T/\rm \varphi}_{\rm H} [p_1,p_2] \equiv \int \frac{d z_1}{2\pi} \frac{d z_2}{2\pi} e^{-i p_1 z_1 -  i p_2 z_2}  \left[ \frac{7}{8} \ln\left(z_1^2\right) + \frac{7}{8} \ln\left(z_2^2\right) + \frac{1}{2} \ln\left((z_1-z_2)^2\right)\right] \notag\\
&\quad \quad \quad \quad \times \left(\theta(2z_s-|z_1|)\theta(z_s-|z_2|)+\theta(z_s-|z_1|)\theta(|z_2|-z_s)\theta(2z_s-|z_2|)\right) \notag\\
&= \frac{7}{8} \left[\text{I0}\left(\left\{2 z_s,-2 z_s\right\},p_2\right) \text{I1}\left(\left\{z_s,-z_s\right\},p_1\right)+\left(\text{I0}\left(\left\{2 z_s,-2
   z_s\right\},p_1\right)-\text{I0}\left(\left\{z_s,-z_s\right\},p_1\right)\right)
   \text{I1}\left(\left\{z_s,-z_s\right\},p_2\right) \right.\notag\\
   &\left.+\text{I0}\left(\left\{z_s,-z_s\right\},p_2\right) \left(\text{I1}\left(\left\{2 z_s,-2
   z_s\right\},p_1\right)-\text{I1}\left(\left\{z_s,-z_s\right\},p_1\right)\right)+\text{I0}\left(\left\{z_s,-z_s\right\},p_1\right) \text{I1}\left(\left\{2
   z_s,-2 z_s\right\},p_2\right)\right] \notag\\
&+\frac{1}{4 \pi  \left(p_1+p_2\right)}\left[ \sin \left(\left(p_1+p_2\right) z_s\right)
   \left[\text{I1}\left(\left\{z_s,-z_s\right\},p_1\right)+\text{I1}\left(\left\{z_s,-z_s\right\},-p_2\right)\right.\right.\notag\\
   &\left.\left.-\text{I1}\left(\left\{2 z_s,-2
   z_s\right\},p_1\right)-\text{I1}\left(\left\{2 z_s,-2 z_s\right\},-p_2\right)+\text{I1}\left(\left\{3 z_s,-3 z_s\right\},p_1\right)+\text{I1}\left(\left\{3
   z_s,-3 z_s\right\},-p_2\right)\right] \right.\notag\\
   &\left.+\sin \left(2 \left(p_1+p_2\right) z_s\right)
   \left[-\text{I1}\left(\left\{z_s,-z_s\right\},p_1\right)-\text{I1}\left(\left\{z_s,-z_s\right\},-p_2\right) \right.\right.\notag\\
   &\left.\left.+\text{I1}\left(\left\{3 z_s,-3
   z_s\right\},p_1\right)+\text{I1}\left(\left\{3 z_s,-3 z_s\right\},-p_2\right)\right] \right.\notag\\
   &\left.+\cos \left(2 \left(p_1+p_2\right) z_s\right)
   \left[-\text{I1t}\left(\left\{z_s,-z_s\right\},p_1\right)+\text{I1t}\left(\left\{z_s,-z_s\right\},-p_2\right) \right.\right.\notag\\
   &\left.\left.+\text{I1t}\left(\left\{3 z_s,-3
   z_s\right\},p_1\right)-\text{I1t}\left(\left\{3 z_s,-3 z_s\right\},-p_2\right)\right] \right.\notag\\
   &\left.+\cos \left(\left(p_1+p_2\right) z_s\right)
   \left[-\text{I1t}\left(\left\{z_s,-z_s\right\},p_1\right)+\text{I1t}\left(\left\{z_s,-z_s\right\},-p_2\right)-\text{I1t}\left(\left\{2 z_s,-2
   z_s\right\},p_1\right)\right.\right.\notag\\
   &\left.\left.+\text{I1t}\left(\left\{2 z_s,-2 z_s\right\},-p_2\right)+\text{I1t}\left(\left\{3 z_s,-3
   z_s\right\},p_1\right)-\text{I1t}\left(\left\{3 z_s,-3 z_s\right\},-p_2\right)\right] \right],
\end{align}
\begin{align}
&I^{T/\rm \varphi}_{\rm HSI} [p_1,p_2] \equiv \int \frac{d z_1}{2\pi} \frac{d z_2}{2\pi} e^{-i p_1 z_1 -  i p_2 z_2} \theta(z_s-|z_1|)\theta(|z_2|-2z_s)\notag
\\&
\quad \quad \quad \quad \, \,\,\times\left[ \frac{7}{8} \ln\left(z_1^2\right) + \frac{7}{8} \ln\left((2 z_s)^2\right) + \frac{1}{2} \ln\left((z_1-2 z_s {\rm sign}[z_2])^2\right)\right] \notag\\
&= \frac{1}{8} \left[4 e^{2 i p_1 z_s} \text{I1}\left(\left\{-z_s,-3 z_s\right\},p_1\right) \text{I0}\left(\left\{\infty ,2 z_s\right\},p_2\right)+4 e^{-2 i p_1
   z_s} \text{I1}\left(\left\{3 z_s,z_s\right\},p_1\right) \text{I0}\left(\left\{-2 z_s,-\infty \right\},p_2\right)\right.\notag\\
   &\left.+7\left(\text{I0}\left(\{\infty ,-\infty
   \},p_2\right)-\text{I0}\left(\left\{2 z_s,-2 z_s\right\},p_2\right)\right) \left(\log \left(4 z_s^2\right)
   \text{I0}\left(\left\{z_s,-z_s\right\},p_1\right)+\text{I1}\left(\left\{z_s,-z_s\right\},p_1\right)\right)\right],
\end{align}
\begin{align}
&I^{T/\rm \varphi}_{\rm HSII}[p_1,p_2] \equiv \int \frac{d z_1}{2\pi} \frac{d z_2}{2\pi} e^{-i p_1 z_1 -  i p_2 z_2} \theta(|z_1|-2z_s)\theta(z_s-|z_2|)\notag
\\&
\quad \quad \quad \quad \quad \times  \left[ \frac{7}{8} \ln\left((2 z_s)^2\right) + \frac{7}{8} \ln\left(z_2^2\right) + \frac{1}{2} \ln\left(({\rm sign}[z_1]2 z_s- z_2)^2\right)\right] \notag\\
&= \frac{1}{8} \left[4 e^{2 i p_2 z_s} \text{I1}\left(\left\{-z_s,-3 z_s\right\},p_2\right) \text{I0}\left(\left\{\infty ,2 z_s\right\},p_1\right)+4 e^{-2 i p_2
   z_s} \text{I1}\left(\left\{3 z_s,z_s\right\},p_2\right) \text{I0}\left(\left\{-2 z_s,-\infty \right\},p_1\right) \right.\notag\\
   &\left.+7\left(\text{I0}\left(\{\infty ,-\infty
   \},p_1\right)-\text{I0}\left(\left\{2 z_s,-2 z_s\right\},p_1\right)\right) \left(\log \left(4 z_s^2\right)
   \text{I0}\left(\left\{z_s,-z_s\right\},p_2\right)+\text{I1}\left(\left\{z_s,-z_s\right\},p_2\right)\right)\right],
\end{align}
\begin{align}
&I^{T/\rm \varphi}_{\rm HSIII} [p_1,p_2] \equiv \int \frac{d z_1}{2\pi} \frac{d z_2}{2\pi} e^{-i p_1 z_1 -  i p_2 z_2} \theta(|z_1|-z_s)\theta(|z_2|-z_s)\theta(z_s-|z_1-z_2|) \notag
\\&
\times \left[ \frac{7}{8} \ln \left(\left(z_s+\left(z_1-z_2\right) \theta \left(z_1-z_2\right)\right){}^2\right)+\frac{7}{8} \ln \left(\left(z_s+\left(z_2-z_1\right)
   \theta \left(z_2-z_1\right)\right){}^2\right) +\frac{1}{2} \ln \left(\left(z_1-z_2\right){}^2\right) \right] 
   \notag\\
&= \frac{1}{8} \text{I0}\left(\{\infty ,-\infty \},p_1+p_2\right) \left[7\log \left(z_s^2\right) \text{I0}\left(\left\{z_s,-z_s\right\},\frac{1}{2}
   \left(p_1-p_2\right)\right)+4\,
   \text{I1}\left(\left\{z_s,-z_s\right\},\frac{1}{2} \left(p_1-p_2\right)\right) \right.
   \notag\\
   &\left. +7e^{\frac{1}{2} i \left(p_1-p_2\right) z_s} \text{I1}\left(\left\{-z_s,-2 z_s\right\},\frac{1}{2} \left(p_1-p_2\right)\right)+7e^{-\frac{1}{2} i \left(p_1-p_2\right) z_s} \text{I1}\left(\left\{2
   z_s,z_s\right\},\frac{1}{2} \left(p_1-p_2\right)\right)\right] 
   \notag\\
   &+\frac{i e^{-i \left(p_1+p_2\right) z_s}}{16 \pi  \left(p_1+p_2\right)} \left[-7\log \left(z_s^2\right) \text{I0}\left(\left\{0,-z_s\right\},p_1\right)+7e^{2 i \left(p_1+p_2\right) z_s} \log
   \left(z_s^2\right) \text{I0}\left(\left\{0,-z_s\right\},-p_2\right) \right.
   \notag\\
   &\left.+7e^{2 i \left(p_1+p_2\right) z_s} \log \left(z_s^2\right)
   \text{I0}\left(\left\{z_s,0\right\},p_1\right)-7\log \left(z_s^2\right) \text{I0}\left(\left\{z_s,0\right\},-p_2\right)-4
   \text{I1}\left(\left\{0,-z_s\right\},p_1\right) \right.
   \notag\\
   &\left.+4 e^{2 i \left(p_1+p_2\right) z_s} \text{I1}\left(\left\{0,-z_s\right\},-p_2\right)-7e^{i p_1 z_s}
   \text{I1}\left(\left\{-z_s,-2 z_s\right\},p_1\right)+7e^{i \left(2 p_1+p_2\right) z_s} \text{I1}\left(\left\{-z_s,-2 z_s\right\},-p_2\right) \right.
   \notag\\
   &\left.+4 e^{2 i\left(p_1+p_2\right) z_s} \text{I1}\left(\left\{z_s,0\right\},p_1\right)-4 \text{I1}\left(\left\{z_s,0\right\},-p_2\right)+7e^{i \left(p_1+2 p_2\right) z_s}
   \text{I1}\left(\left\{2 z_s,z_s\right\},p_1\right)-7e^{i p_2 z_s} \text{I1}\left(\left\{2 z_s,z_s\right\},-p_2\right)\right],
\end{align}
\begin{align}
&I^{T/\rm \varphi}_{\rm HSIV} [p_1,p_2] \equiv \int \frac{d z_1}{2\pi} \frac{d z_2}{2\pi} e^{-i p_1 z_1 -  i p_2 z_2} \theta(|z_1|-z_s)\theta(|z_2|-z_s)\theta(z_s-|z_1+z_2|) \notag
\\&
\times \left[ \frac{7}{8} \ln \left(\left(z_s+\left(z_1+z_2\right) \theta \left(z_1+z_2\right)\right){}^2\right)+\frac{7}{8} \ln \left(\left(-z_s+\left(z_2+z_1\right)
   \theta \left(-z_2-z_1\right)\right){}^2\right)+\frac{1}{2} \ln \left(\left(2 z_s + |z_1+z_2|\right){}^2\right) \right]  
   \notag\\
   &=\frac{1}{16} \text{I0}\left(\{\infty ,-\infty \},\frac{1}{2} \left(p_1-p_2\right)\right) \left[7\log \left(z_s^2\right)
   \text{I0}\left(\left\{z_s,-z_s\right\},\frac{1}{2} \left(p_1+p_2\right)\right)\right.
   \notag\\
   &\left.+4 e^{i \left(p_1+p_2\right) z_s} \text{I1}\left(\left\{-2 z_s,-3z_s\right\},\frac{1}{2} \left(p_1+p_2\right)\right) +7e^{\frac{1}{2} i \left(p_1+p_2\right) z_s} \text{I1}\left(\left\{-z_s,-2 z_s\right\},\frac{1}{2}
   \left(p_1+p_2\right)\right)\right.
   \notag\\
   &\left.+7e^{-\frac{1}{2} i \left(p_1+p_2\right) z_s} \text{I1}\left(\left\{2 z_s,z_s\right\},\frac{1}{2} \left(p_1+p_2\right)\right)+4
   e^{-i \left(p_1+p_2\right) z_s} \text{I1}\left(\left\{3 z_s,2 z_s\right\},\frac{1}{2} \left(p_1+p_2\right)\right)\right] 
   \notag\\
   &+\frac{i e^{-i \left(p_1+p_2\right) z_s}}{16 \pi 
   \left(p_1-p_2\right)} \left[-7e^{2 i p_2 z_s} \log \left(z_s^2\right) \text{I0}\left(\left\{0,-z_s\right\},p_1\right)+7e^{2 i p_1 z_s} \log
   \left(z_s^2\right) \text{I0}\left(\left\{0,-z_s\right\},p_2\right)\right.
   \notag\\
   &\left.+7e^{2 i p_1 z_s} \log \left(z_s^2\right)
   \text{I0}\left(\left\{z_s,0\right\},p_1\right)-7e^{2 i p_2 z_s} \log \left(z_s^2\right) \text{I0}\left(\left\{z_s,0\right\},p_2\right)-4 e^{2 i\left(p_1+p_2\right) z_s} \text{I1}\left(\left\{-2 z_s,-3 z_s\right\},p_1\right)\right.
   \notag\\
   &\left.+4 e^{2 i \left(p_1+p_2\right) z_s} \text{I1}\left(\left\{-2 z_s,-3z_s\right\},p_2\right)-7e^{i \left(p_1+2 p_2\right) z_s} \text{I1}\left(\left\{-z_s,-2 z_s\right\},p_1\right)+7e^{i \left(2 p_1+p_2\right) z_s}
   \text{I1}\left(\left\{-z_s,-2 z_s\right\},p_2\right)\right.
   \notag\\
   &\left.+7e^{i p_1 z_s} \text{I1}\left(\left\{2 z_s,z_s\right\},p_1\right)-7e^{i p_2 z_s} \text{I1}\left(\left\{2
   z_s,z_s\right\},p_2\right)+4 \text{I1}\left(\left\{3 z_s,2 z_s\right\},p_1\right)-4 \text{I1}\left(\left\{3 z_s,2 z_s\right\},p_2\right)\right],
\end{align}
\begin{align}
&I^{T/\rm \varphi}_{\rm S} [p_1,p_2] \equiv \int \frac{d z_1}{2\pi} \frac{d z_2}{2\pi} e^{-i p_1 z_1 -  i p_2 z_2}  \theta(|z_1|-z_s)\theta(|z_2|-z_s)\theta(|z_1-z_2|-z_s)\theta(|z_1+z_2|-z_s) \notag
\\
&
\quad \quad \quad \quad \times\left[ \frac{7}{8} \ln\left(z_s^2\right) + \frac{7}{8} \ln\left((2z_s)^2\right) + \frac{1}{2} \ln\left(({\rm sign}[z_1]z_s-{\rm sign}[z_2]2z_s)^2\right)\right]
\notag\\
&=\delta \left(p_1-p_2\right) \delta \left(p_1+p_2\right) \left(\frac{7}{4} \log \left(4 z_s^4\right)+\frac{1}{2} \log \left(9 z_s^4\right)\right) 
\notag\\
& -\frac{\delta \left(p_1+p_2\right) \left(4 \log \left(z_s^2\right)+7 \log \left(4 z_s^4\right)\right) \sin \left(\frac{1}{2} \left(p_1-p_2\right) z_s\right)}{4
   \pi  \left(p_1-p_2\right)}
   \notag\\
&-\frac{\delta \left(p_1-p_2\right) \left(4 \log \left(9 z_s^2\right)+7 \log \left(4 z_s^4\right)\right) \sin \left(\frac{1}{2}
   \left(p_1+p_2\right) z_s\right)}{4 \pi  \left(p_1+p_2\right)} 
   \notag\\
& -\frac{\delta \left(p_2\right) \left(18 \log \left(z_s\right)+\log (1152)\right) \sin \left(p_1 z_s\right)}{4 \pi  p_1}-\frac{\delta \left(p_1\right) \left(18
   \log \left(z_s\right)+\log (1152)\right) \sin \left(p_2 z_s\right)}{4 \pi  p_2}
   \notag\\
&-\frac{\left(18 \log \left(z_s\right)+\log (128)\right) \left(p_1 \cos \left(\left(p_1+2 p_2\right) z_s\right)+p_2 \cos \left(\left(2 p_1+p_2\right)
   z_s\right)\right)}{8 \pi ^2 p_1 p_2 \left(p_1+p_2\right)}
   \notag\\
&+\frac{\left(18 \log \left(z_s\right)+\log (10368)\right) \left(p_1 \cos \left(\left(p_1-2 p_2\right) z_s\right)-p_2 \cos \left(2 p_1 z_s-p_2 z_s\right)\right)}{8 \pi ^2 p_1 \left(p_1-p_2\right) p_2}.
\end{align}

\end{document}